\documentclass[iop]{emulateapj}
\usepackage{natbib}
\usepackage{graphicx}
\usepackage{hyperref}
\usepackage{breakurl}
\usepackage{apjfonts}
\newcommand{\msun} {$M_{\odot}$}

\newcommand{\rsun} {$R_{\odot}$}

\newcommand{\hbeta} {H$\beta$}
\newcommand{\hgamma} {H$\gamma$}
\newcommand{\hdelta} {H$\delta$}
\newcommand{\hepsilon} {H$\epsilon$}
\newcommand{\kms} {km s$^{-1}$}
\newcommand{\Te} {$T_{\rm eff}$}
\newcommand{\logg} {$\log g$}
\newcommand{\loghe} {$\log$~(He/H)}
\newcommand{\logca} {$\log$~(Ca/H)}
\newcommand{\logmg} {$\log$~(Mg/H)}
\newcommand{\heii} {He {\sc ii} $\lambda$4471}
\newcommand{\heib} {He {\sc i} $\lambda$4026}
\newcommand{\heia} {He {\sc i} $\lambda$4471}

\newcommand{\mgii} {Mg {\sc ii} $\lambda$4481}
\newcommand{\caiih} {Ca {\sc ii} $\lambda$3968}
\newcommand{\caiik} {Ca {\sc ii} $\lambda$3934}
\newcommand{\psr} {PSR~J1816+4510}
\newcommand{\lp} {LP~400-22}
\newcommand{\nltt} {NLTT~11748}
\newcommand{\galex} {GALEX~J1717+6757}

\begin{document}

\slugcomment{Accepted for publication in ApJ}
\shortauthors{GIANNINAS ET AL}
\shorttitle{ATMOSPHERIC PARAMETERS OF SHORTEST PERIOD BINARY WDs}

\title{PRECISE ATMOSPHERIC PARAMETERS FOR THE SHORTEST PERIOD BINARY
  WHITE DWARFS: GRAVITATIONAL WAVES, METALS, AND PULSATIONS*}


\author{A. Gianninas$^{1}$, P. Dufour$^{2}$, Mukremin Kilic $^{1}$, 
	Warren R. Brown$^{3}$, P. Bergeron$^{2}$ and J.J. Hermes$^{4}$}

\affil{$^{1}$Homer L. Dodge Department of Physics and Astronomy,
  University of Oklahoma, 440~W.~Brooks~St., Norman, OK 73019, USA;
  alexg@nhn.ou.edu}
\affil{$^{2}$D\'epartement de Physique, Universit\'e de Montr\'eal,
  C.P.~6128, Succ.~Centre-Ville, Montr\'eal, Qu\'ebec H3C 3J7, Canada}
\affil{$^{3}$Smithsonian Astrophysical Observatory, 60~Garden~St.,
  Cambridge, MA 02138, USA}
\affil{$^{4}$Department of Physics, University of Warwick, Coventry
  CV4 7AL, UK}

\begin{abstract}

We present a detailed spectroscopic analysis of 61 low mass white
dwarfs and provide precise atmospheric parameters, masses, and updated
binary system parameters based on our new model atmosphere grids and
the most recent evolutionary model calculations.  For the first time,
we measure systematic abundances of He, Ca and Mg for metal-rich
extremely low mass white dwarfs and examine the distribution of these
abundances as a function of effective temperature and mass. Based on
our preliminary results, we discuss the possibility that shell flashes
may be responsible for the presence of the observed He and
metals. We compare stellar radii derived from our spectroscopic
analysis to model-independent measurements and find good agreement
except for those white dwarfs with \Te\ $\lesssim$~10,000~K. We
also calculate the expected gravitational wave strain for each system
and discuss their significance to the $eLISA$ space-borne
gravitational wave observatory. Finally, we provide an update on the
instability strip of extremely low mass white dwarf pulsators.

\end{abstract}

\keywords{binaries: close -- stars: abundances -- stars: fundamental
  parameters -- techniques: spectroscopic -- white dwarfs}

\footnotetext[*]{Based on observations obtained at the MMT
  Observatory, a joint facility of the Smithsonian Institution and the
  University of Arizona.}

\section{INTRODUCTION}

Extremely low mass (ELM) white dwarfs (WDs) with surface gravities of
\logg~$\lesssim$~7.0 (or masses $M$~$\lesssim$~0.30~\msun) are
presumed to have He cores and are necessarily the product of the
evolution of compact binary systems. The Universe is not yet old
enough to have produced such ELM WDs through normal single-star
evolution \citep{marsh95}. These extreme products of binary evolution
represent the possible progenitors of type Ia supernovae
\citep{iben84}, underluminous .Ia supernovae \citep{bildsten07}, AM
CVn systems \citep{breedt12,kilic14b} and possibly even R~CrB stars 
\citep{webbink84,clayton13}.

One of the first spectroscopically confirmed ELM WDs was found as
the companion to the millisecond pulsar PSR J1012+5307 
\citep{vk96,callanan98}. Several more ELM WDs have been 
spectroscopically identified as pulsar companions \citep[e.g. ][ and
references therein]{antoniadis13,kaplan13,kaplan14a,ransom14,smedley14} 
and in other short period binary systems
\citep{heber03,liebert04,kawka06,mullally09,kulkarni10,marsh11,vennes11,silvotti12}.
  
Based on a comparison of the mass distribution of post-common
envelope binaries and wide WD+main sequence binaries from the Sloan
Digital Sky Survey (SDSS), \citet{rm11} confirmed that the majority
of low-mass WDs reside in close binary systems \citep{marsh95}.
Furthermore, the EL CVn-type binaries, with orbital periods 
$P_{\rm orb}$ $\approx$~0.7--2.2 d, published in \citet{maxted13} and
\citet{maxted14} also represent potential progenitors to ELM WDs.

Existing in such tight binary systems, we expect ELM WDs to be sources
of gravitational waves \citep{hermes12c,kilic13a} as their orbits
decay due to the loss of orbital angular momentum. Hence, they
represent potential testbeds for general relativity and the shortest
period systems serve as verification sources for future gravitational
wave detectors such as $eLISA$ \citep{amaro13}. The close nature of
these systems also gives rise to phenomena such as ellipsoidal
variations due to tidal distortions and Doppler beaming. Both of these
phenomena manifest themselves in the light curves of ELM WDs. The
analysis of ellipsoidal variations \citep{hermes12a,gianninas14}, as
well as parallax measurements \citep{kilic13b} and eclipse modeling
\citep{hermes12c,kaplan14b,bours14}, provide model-independent methods
for measuring the stellar radius. Recent studies
\citep{kilic13b,kaplan14b,gianninas14} have brought to light a
discrepancy between the radii inferred from spectroscopic analyses to
those measured by model-independent methods for the coolest ELM WDs.
Precise measurements of the atmospheric parameters of ELM WDs are
needed in order to shed light on this issue.
 
The ELM Survey
\citep{brown_ELM1,brown_ELM3,brown_ELM5,kilic_ELM0,kilic_ELM2,kilic_ELM4}
has been searching for short-period ($P_{\rm orb} \leqslant$~1~day)
ELM WD binaries for several years now with considerable success. 
Candidates are selected mostly using available SDSS photometry and
then followed-up with optical time-series spectroscopy through
which radial velocity (RV) variations are detected. In some
cases, SDSS spectroscopy has also been useful in identifying
potential ELM WD candidates. So far, over 60 ELM WD binaries have
been discovered by the ELM Survey and over 30 of them will merge
within a Hubble time \citep{brown_ELM5}.
 
All of the previously published ELM Survey analyses have noted
the presence of the \caiik\ K resonance line in the spectra of all
ELM WDs with \logg~$\leqslant$~6.0. The fact that all of the
lowest mass ELM WDs share this spectroscopic signature suggests that
a physical phenomenon related to their evolution might be
responsible. However, a systematic study of the metal abundances in
ELM WDs has never been performed; abundances have only been measured
for a handful of systems. These include the hot ELM WD
\galex\ \citep{vennes11,hermes14b}, the WD companion to
\psr\ \citep{kaplan13}, and, most recently, \citet{gianninas14}
analyzed the unusually metal-rich and tidally distorted ELM WD binary
J0745+1949 (hereafter, J0745).

For canonical mass WDs, \citet{zuckerman03,zuckerman10} showed
that $\approx$~25\% of hydrogen atmosphere DA WDs and $\approx$~30\%
of helium atmosphere DB WDs are polluted with metals based on a
detailed analysis of high resolution spectroscopy. The more
recent analysis of \citet{koester14} suggests that the number of
metal-rich WDs is closer to 60\%. For WDs with \Te\ $\leq$~20,000~K, 
this phenomenon is understood to be the consequence of ongoing 
accretion from a circumstellar disk resulting
from the tidal disruption of a rocky body venturing too close to the
host WD. This has been confirmed through the detection of infrared
(IR) excesses due to the emission of a dusty
\citep[e.g.][]{jura03,kilic06,farihi09,barber12} or gaseous
\citep[e.g.][and references therein]{melis12,gansicke06,gansicke11} 
debris disk surrounding the WD. To date, there is no evidence that 
analogous disks are present around polluted ELM WD binaries.

\citet{kaplan13} suggested that the observed metals could be the
result of a recent shell flash that would serve to mix the outer
layers of the WD and bring metals to the surface. The evolutionary 
models of \citet{althaus13} certainly suggest that ELM WDs with masses 
between 0.18~\msun\ and 0.36~\msun\ undergo a series of H-shell flashes 
as they evolve. However, this would not explain the presence of Ca 
for ELM WDs with masses $<$~0.18~\msun, where shell flashes are not 
predicted.

In this paper, we present a comprehensive and homogeneous
spectroscopic analysis of the entire ELM Survey sample including
updated measurements of the atmospheric parameters, \Te\ and \logg, as
well as abundances of all the observed metals. We also provide
improved mass estimates for both components of the system and use
these to calculate the expected gravitational wave strain of the
system. Furthermore, this is the first time we have a large
enough sample of ELM WDs with independent radius measurements to
explore the discrepancies with spectroscopically derived radii. We
summarize, in Section 2, the current sample of WDs from the ELM Survey
and briefly describe our observations. Section 3 lists and explains
the various grids of model atmospheres used in our study. In Section
4, we present the results of our analysis including our updated
physical and binary parameters.  Section 5 discusses the ensemble
properties of our sample.  Finally, Section 6 outlines our conclusions
and we comment on future avenues of research.

\section{ELM SURVEY SAMPLE}

The sample that we analyze includes a total of 61 ELM WD binaries from
the ELM Survey. The bulk of this sample is comprised of the 58 ELM WDs
listed in Table 3 of \citet{brown_ELM5} but also includes three
additional ELM WDs which have been published in separate papers since
then. These three ELM WDs are the metal-rich and tidally distorted ELM
WD J0745 \citep{gianninas14} and the two pulsating ELM WDs J1614+1912
and J2228+3623 \citep{hermes13c}. Of these 61 ELM WDs, 55 are
confirmed as being in short-period binary systems through the
detection of RV variability and their orbital periods ($P$) and
velocity semi-amplitudes ($K$) are well constrained. The six remaining
ELM WDs (J0900+0234, J1448+1342, J1614+1912, J2228+3623, J2252$-$0056,
J2345$-$0102) do not display significant radial velocity variability
and only upper limits for $K$ have been measured.

The spectra of these 61 ELM WDs were obtained using five distinct
setups on two different telescopes. A total of 57 targets were
observed at the 6.5m MMT telescope with the Blue Channel spectrograph
\citep{schmidt89}. With only one exception, the observations were
obtained using the 832~line~mm$^{-1}$ grating but with two different
slit widths. First, 35 targets were observed using a 1$\farcs$0 slit
providing a spectral resolution of 1.0~\AA\ and an additional 21
targets were observed using a 1$\farcs$25 slit achieving a spectral
resolution of 1.2~\AA. Finally, the spectra of J0651+2844 (hereafter,
J0651) were obtained using the 800~line~mm$^{-1}$ grating coupled with
a 1$\farcs$0 slit producing a resolution of 2.3~\AA. All the spectra
obtained at the MMT provide spectral coverage from 3600~\AA\ to
4500~\AA\ spanning the Balmer series from \hgamma\ to the Balmer jump.

The four remaining targets were observed using the Fred Lawrence
Whipple Observatory's (FLWO) 1.5m Tilinghast telescope equipped with
the FAST spectrograph \citep{fabricant98} and the 600~line~mm$^{-1}$
grating. Three targets were observed with a 2$\farcs$0 slit providing
a resolution of 2.3~\AA\ with the remaining target having been
observed through a 1$\farcs$5 slit for a resolution of 1.7~\AA. The
observations obtained at FLWO provide a slightly better spectral
coverage than the MMT observations, covering from 3500~\AA\ to
5500~\AA\ and thus include \hbeta\ as well.

We note that the nature of the ELM Survey and the necessity of
obtaining multiple observations for each ELM WD system to confirm its
RV variability and, subsequently, to sample the full binary orbit,
provides high signal-to-noise ratio (S/N) observations for the
majority of the ELM WDs in our sample with the exception of the very
faintest targets. High S/N observations are crucial if we hope to
accurately and precisely determine the atmospheric parameters of ELM
WDs \citep[see Section 3 and Fig. 12 of][ for a demonstration of the
importance of S/N in the determination of \Te\ and \logg]{gianninas05}.

Finally, the excellent data quality has also allowed us to easily
discern the presence of Ca, as well as Mg, in the atmosphere of many
of these ELM WDs. There have been cases where observed Ca lines have
been identified as being interstellar in origin. Notably,
\citet{silvotti12}, who analyzed the sdB+WD system KIC~6614501, and
\citet{kaplan14a}, in their study of one of the WD companions to
PSR~J0337+1715, concluded that the Ca lines they observed in the WD
spectrum were interstellar in origin. This conclusion was reached by
observing that the Ca lines were not Doppler-shifted along with the
Balmer lines. In the ELM Survey sample, the individual spectra used
for the RV measurements tend to show both a stationary interstellar
component and a photospheric component whose RV correlates with that
of the Balmer lines. In all cases, the photospheric component
dominates but only a few WDs have enough spectra obtained at
quadrature, and at a high enough S/N, to even attempt to separate the
two components.  We therefore caution that our Ca abundances should be
considered as upper limits.

\newpage

\section{MODEL ATMOSPHERES}

\subsection{Pure Hydrogen Model Atmospheres}

For the analysis of the hydrogen Balmer lines, we use hydrogen-rich
model atmospheres and synthetic spectra that are derived from the
model atmosphere code originally described in \citet{bergeron95} and
references therein, with recent improvements discussed in
\citet{TB09}. Briefly, our models assume a plane-parallel geometry,
hydrostatic equilibrium and local thermodynamic equilibrium (LTE). The
assumption of LTE is justified as our model grid is restricted to
\Te\ $\leqslant$ 35,000~K where NLTE effects are not yet significant,
even for \logg\ $<$~7.0 \citep{napiwotzki97}. Furthermore, our models
adopt the ML2/$\alpha$~=~0.8 parametrization of the mixing length
theory as prescribed by \citet{tremblay10}. Finally, we utilize the
new Stark broadening profiles from \citet{TB09} that include the
occupation probability formalism of \citet{hm88} directly in the line
profile calculation. For the purposes of fitting the spectra of our
ELM WD sample, we have computed a new model grid which we have
extended to much lower surface gravities. Our full model grid covers
\Te\ from 4000 K to 35,000~K in steps ranging from 250 to 5000~K, and
\logg\ from 4.5 to 9.5 in steps of 0.25 dex.

\subsection{Mixed Hydrogen/Helium Model Atmospheres}

For the five ELM WDs which contain helium lines in their optical
spectra, we have computed a separate grid of models. This distinct
model grid covers \Te\ from 4500~K to 30,000~K in steps ranging from
250 to 5000~K, \logg\ from 4.75 to 8.0 in steps of 0.25 dex and
\loghe\ from $-$4.0 to 0.0 in steps of 1.0 dex. These models are
identical to the pure hydrogen models but also include helium which is
treated according to the formalism presented in \citet{bergeron11},
including the improved Stark broadening profiles from
\citet{beauchamp97} for over 20 lines of neutral helium. These line
profiles are similar to those presented in \citet{beauchamp96} with
the exception that at low temperatures (\Te~$<$~10,800~K) the
free-free absorption coefficient of the negative helium ion of
\citet{john94} is now used.

\subsection{Model Atmospheres with Metals}

To measure the abundances of Mg \& Ca based on their observed
absorption lines, we computed separate grids of model atmospheres and
synthetic spectra. We performed these calculations using the same code
that was used to model the heavily metal polluted DBZ star J0738+1835
\citep{dufour12} and the metal-rich ELM WD J0745
\citep{gianninas14}. Keeping \Te\ ad \logg\ fixed at the values
determined from the Balmer line fits of each metal-rich ELM WD, we
proceed to calculate several grids of synthetic spectra, one for each
element of interest (i.e. Mg \& Ca). The individual grids cover a
range of abundances from $\log[n({\rm Z})/n({\rm H})]$~=~$-$3.0 to
$-$10.0, in steps of 0.5~dex.

\begin{figure*}[!hp]
\begin{minipage}[c][][t]{\textwidth}
\centering
\includegraphics[scale=0.585,angle=-90,bb=40 57 592 739]{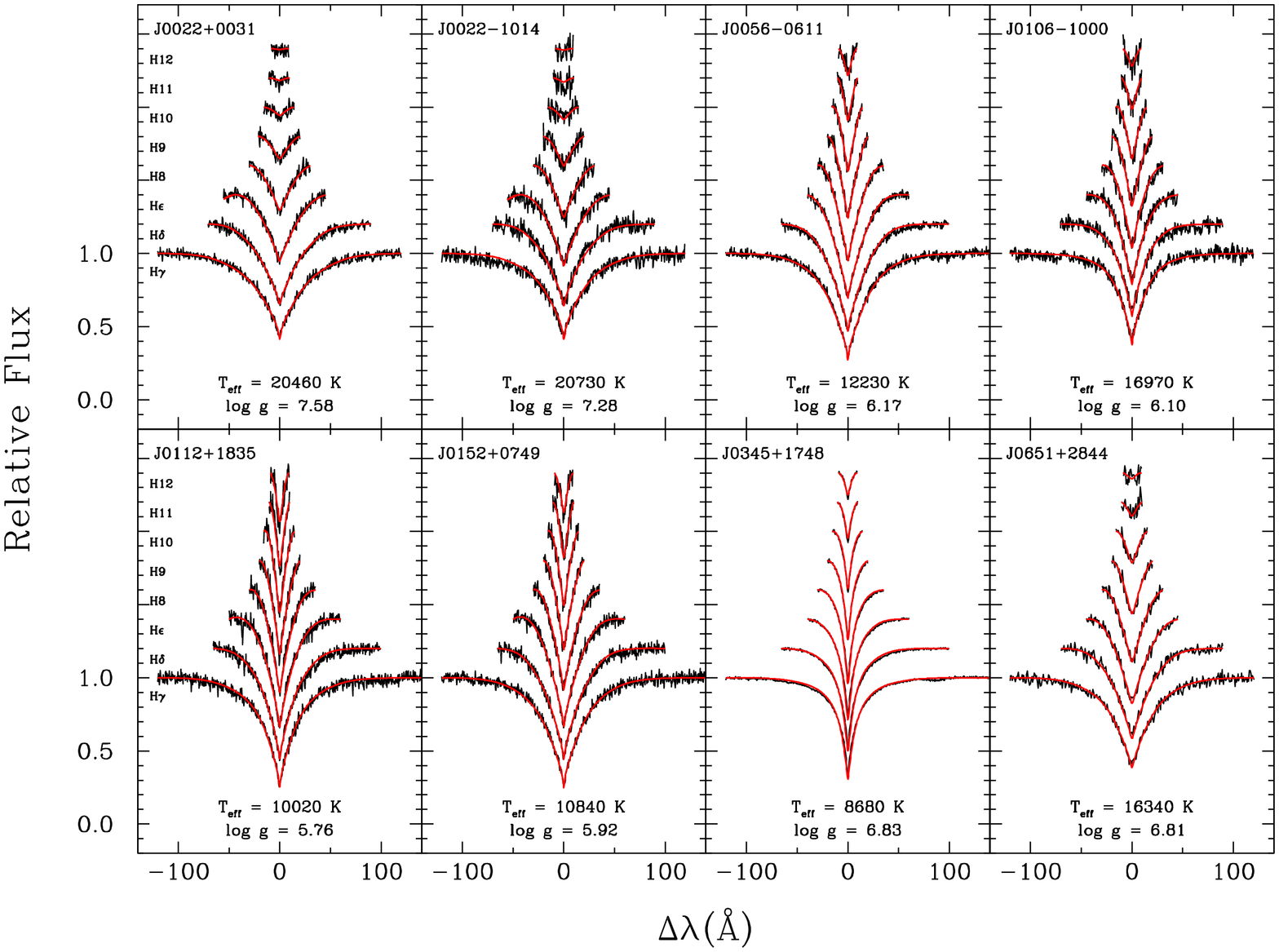}
\par\vfill
\includegraphics[scale=0.585,angle=-90,bb=40 57 592 739]{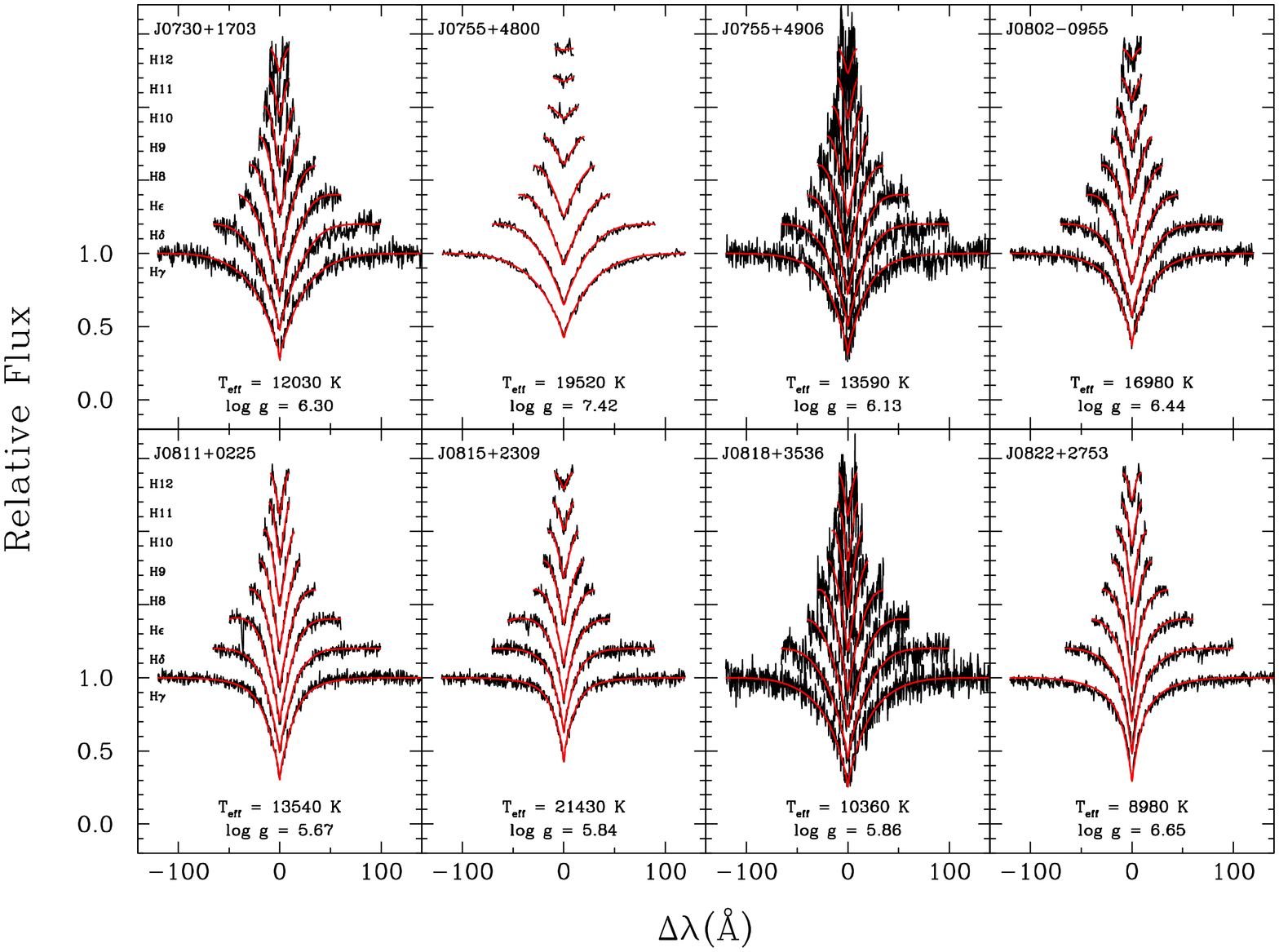}
\figcaption[f01a.eps]{Model fits (red) to the individual Balmer
  line profiles (black) for 55 WDs from the ELM Survey. The lines
  range from \hgamma\ (bottom) to H12 (top), each offset by a factor
  of 0.2 for clarity. The best-fit values of \Te\ and \logg\ are
  indicated at the bottom of each panel.  \label{fg:DA1}}
\end{minipage}
\end{figure*}

\addtocounter{figure}{-1}
\begin{figure*}[!hp]
\begin{minipage}[c][][t]{\textwidth}
\centering
\includegraphics[scale=0.585,angle=-90,bb=40 57 592 739]{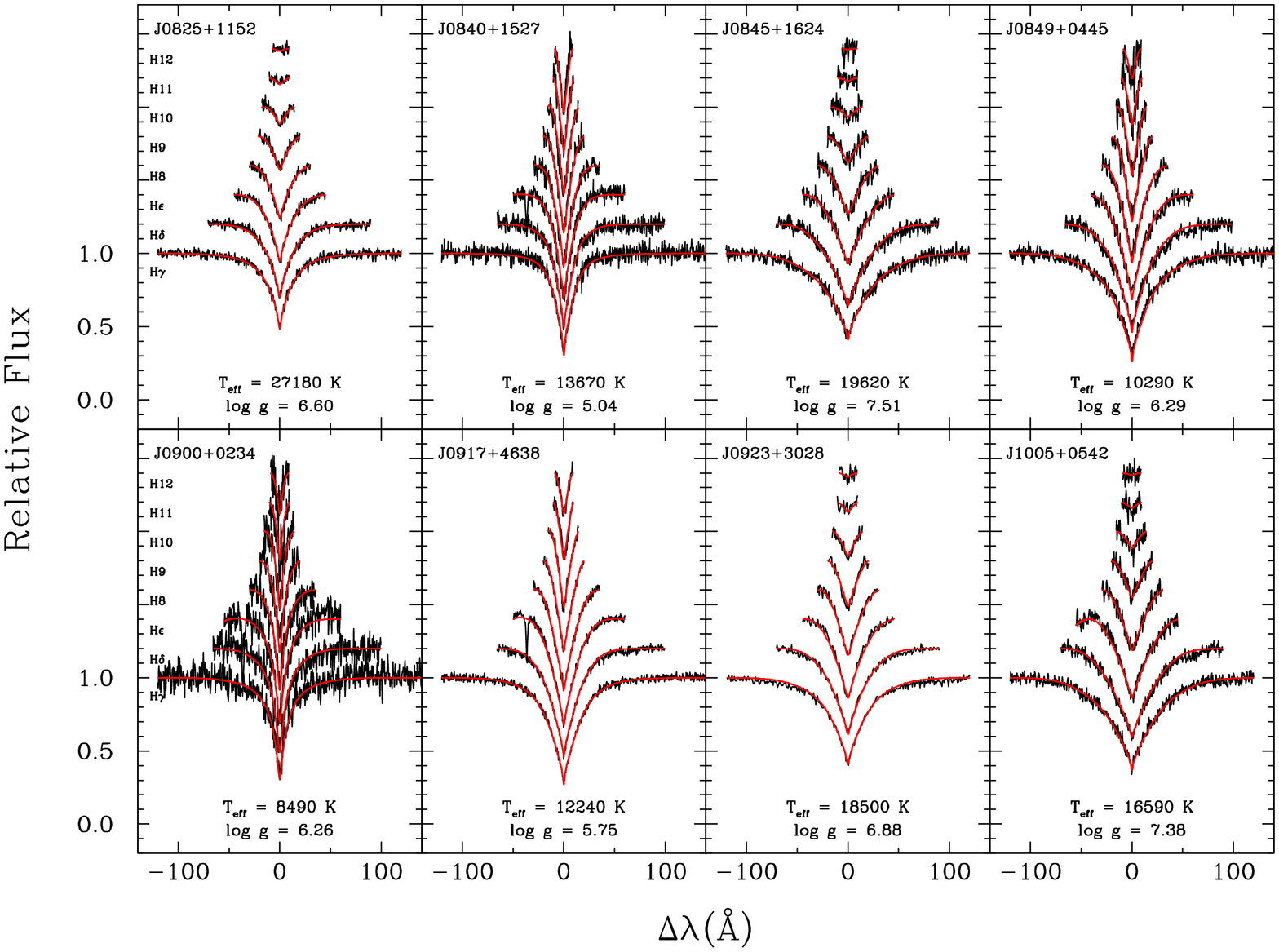}
\par\vfill
\includegraphics[scale=0.585,angle=-90,bb=40 57 592 739]{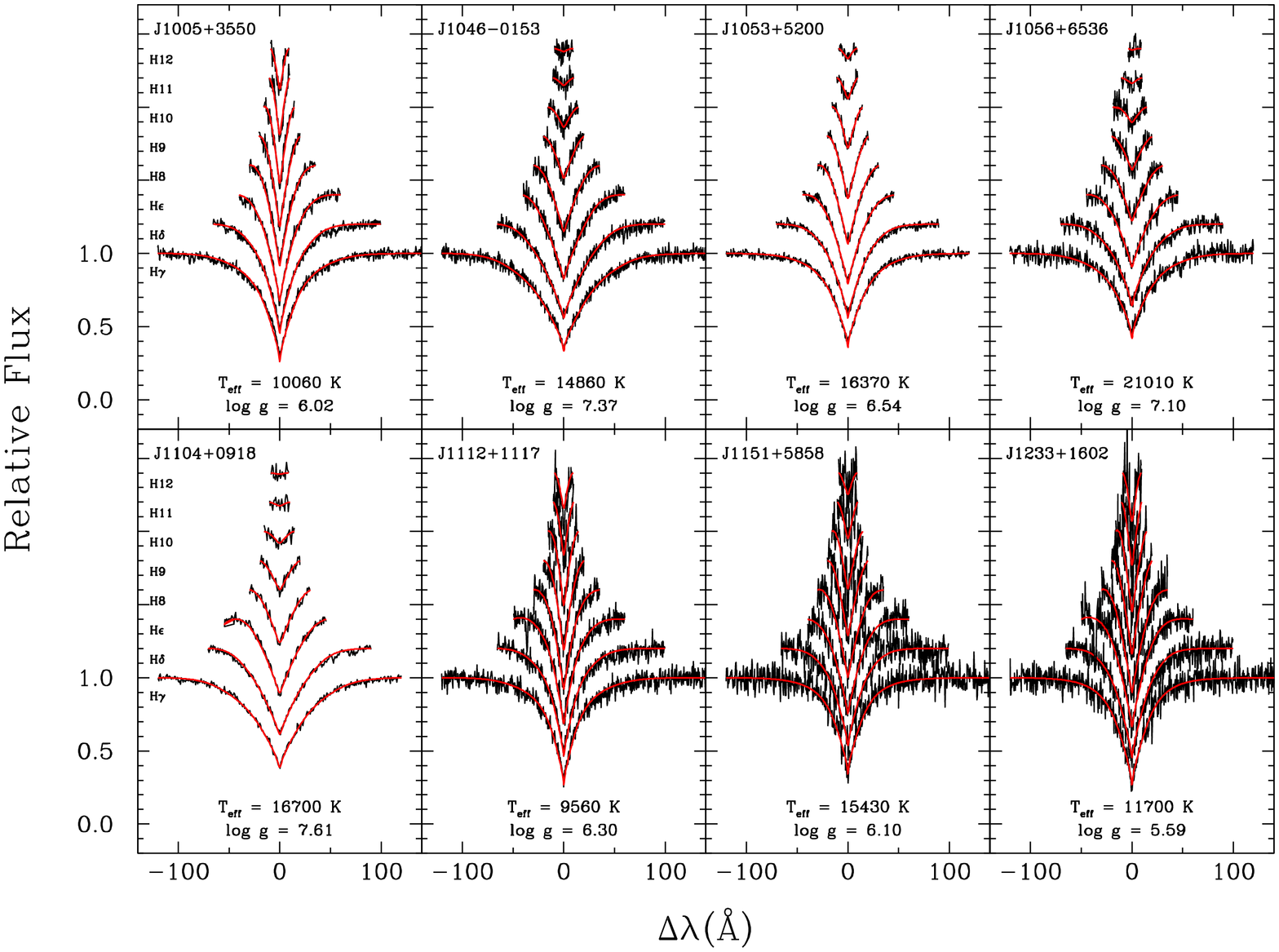}
\figcaption[f01c.eps]{(Continued) \label{fg:DA2}}
\end{minipage}
\end{figure*}

\addtocounter{figure}{-1}
\begin{figure*}[!hp]
\begin{minipage}[c][][t]{\textwidth}
\centering
\includegraphics[scale=0.585,angle=-90,bb=40 57 592 739]{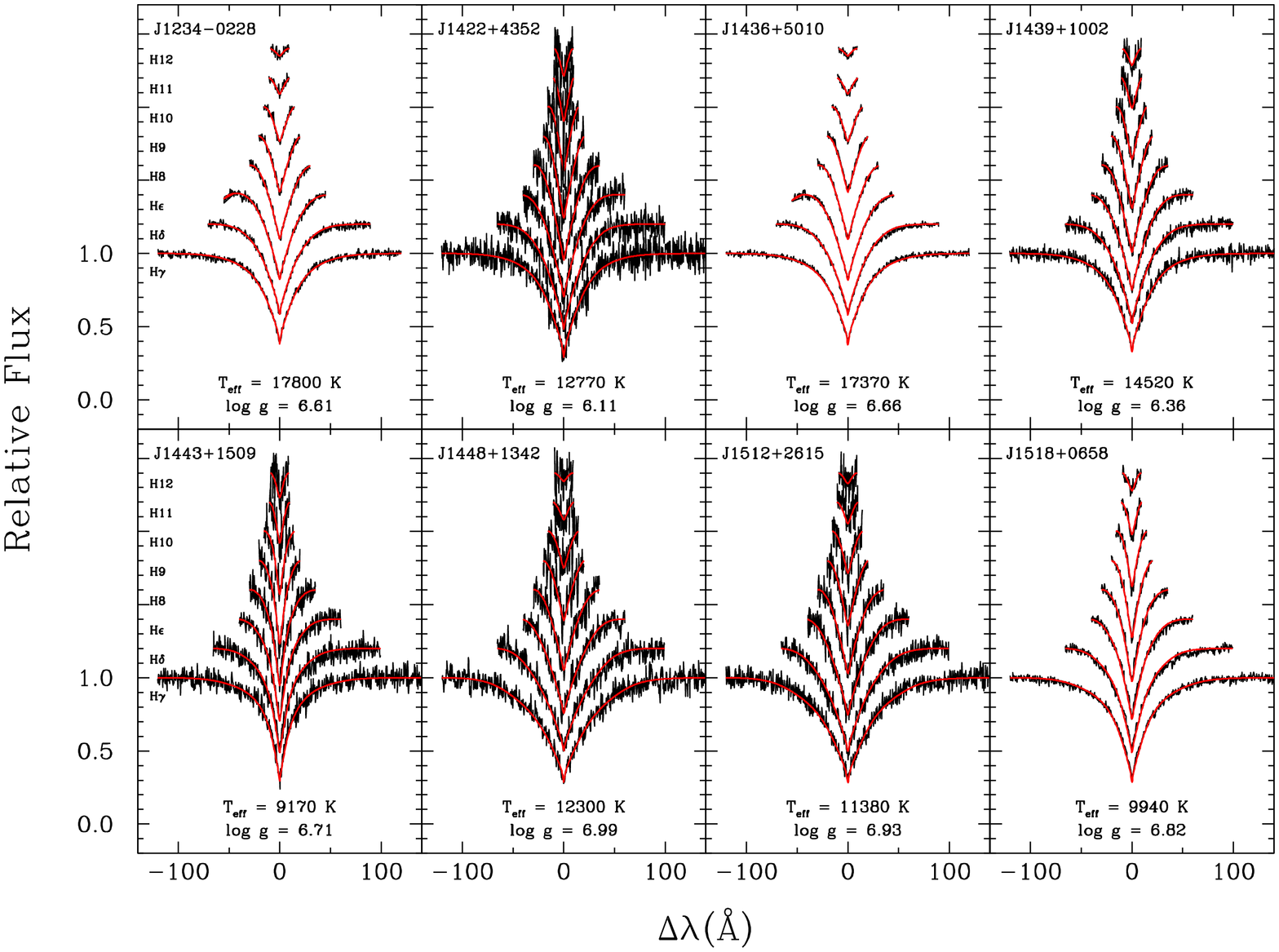}
\par\vfill
\includegraphics[scale=0.585,angle=-90,bb=40 57 592 739]{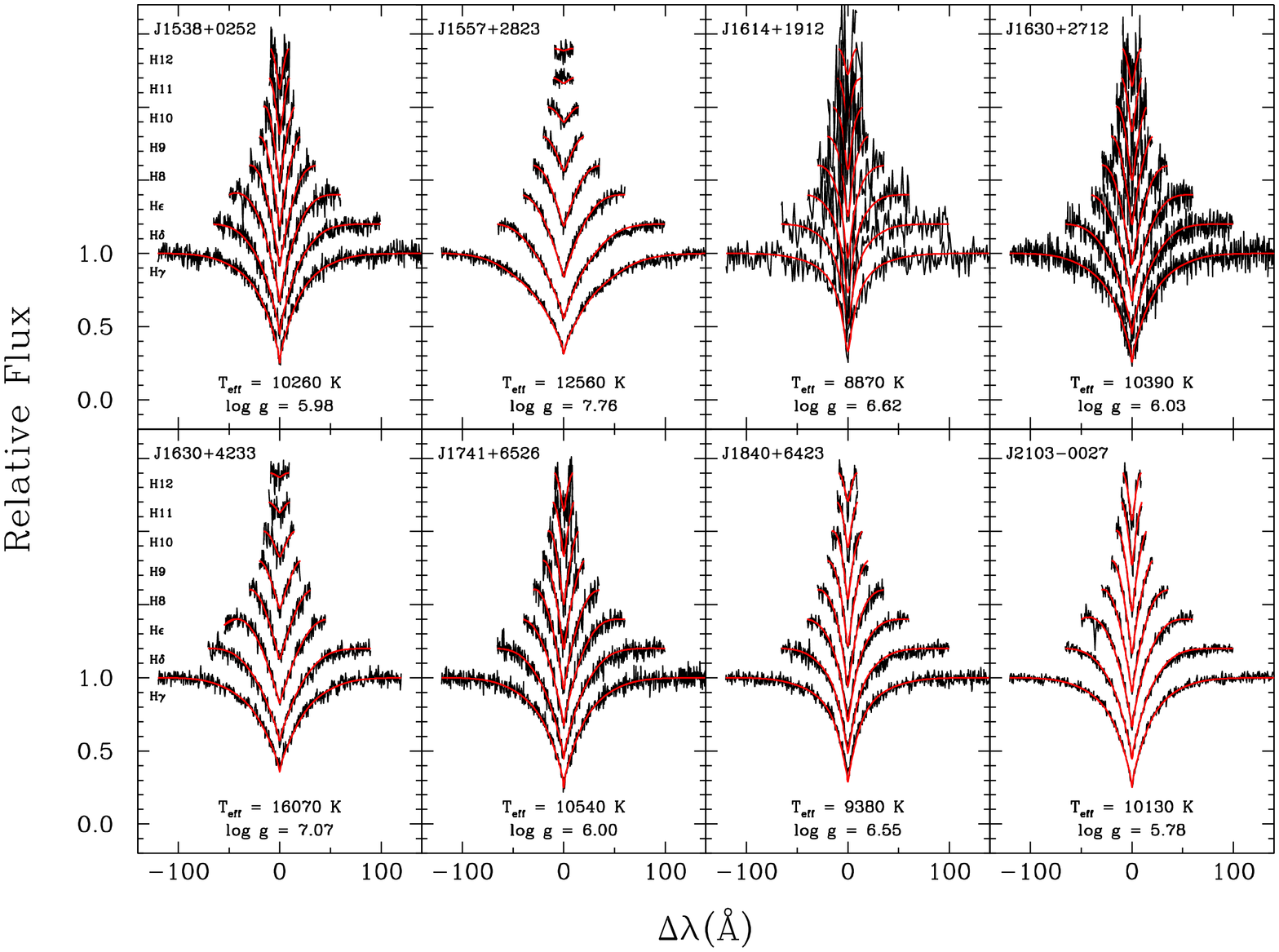}
\figcaption[f01f.eps]{(Continued) \label{fg:DA3}}
\end{minipage}
\end{figure*}

\addtocounter{figure}{-1}
\begin{figure*}[!h]
\begin{minipage}[c][][t]{\textwidth}
\centering
\includegraphics[scale=0.585,angle=-90,bb=40 57 592 739]{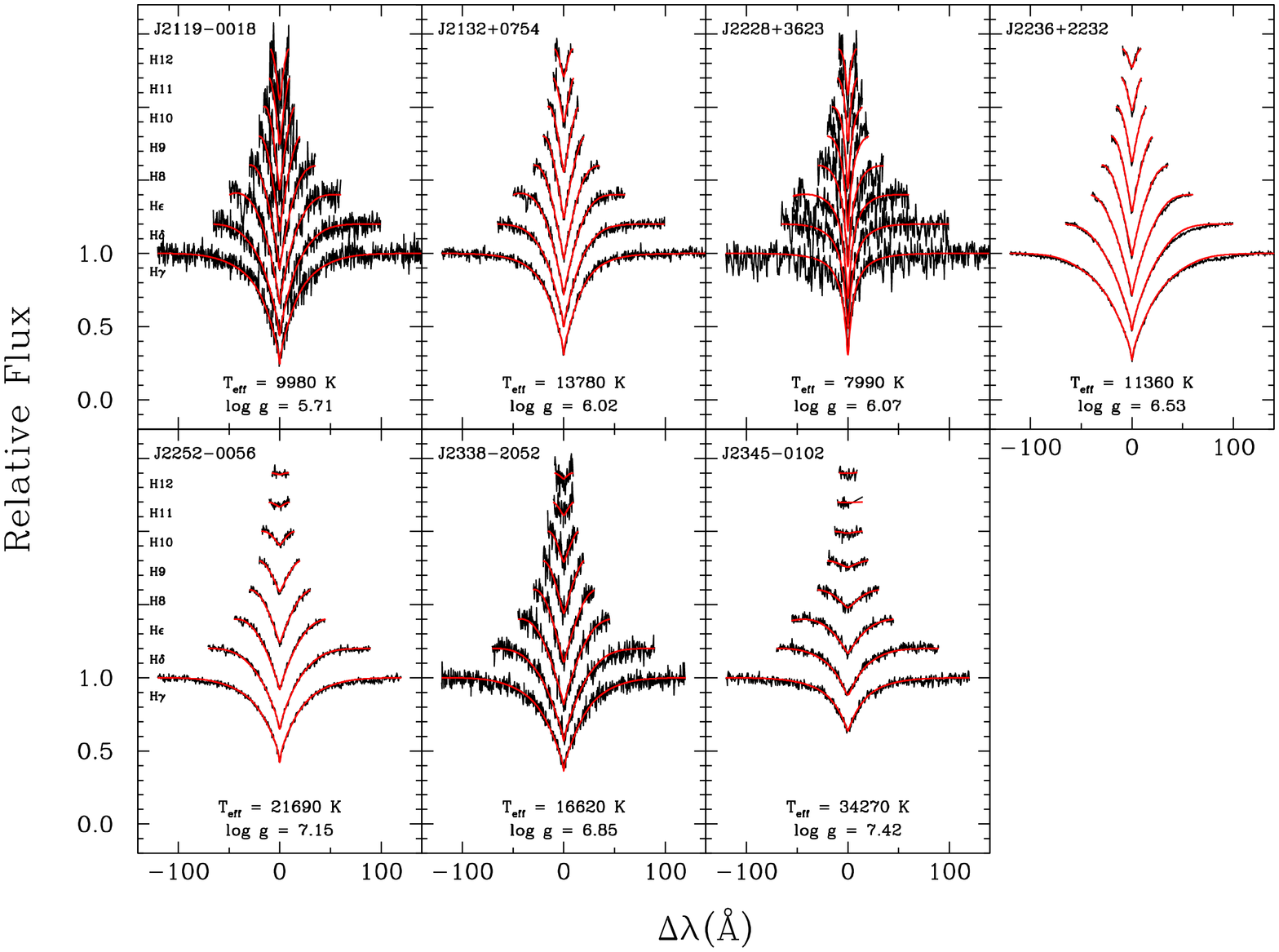}
\figcaption[f01g.eps]{(Continued) \label{fg:DA4}}
\end{minipage}
\end{figure*}

\begin{figure*}[!h]
\includegraphics[scale=0.665,angle=-90,bb=122 9 526 784]{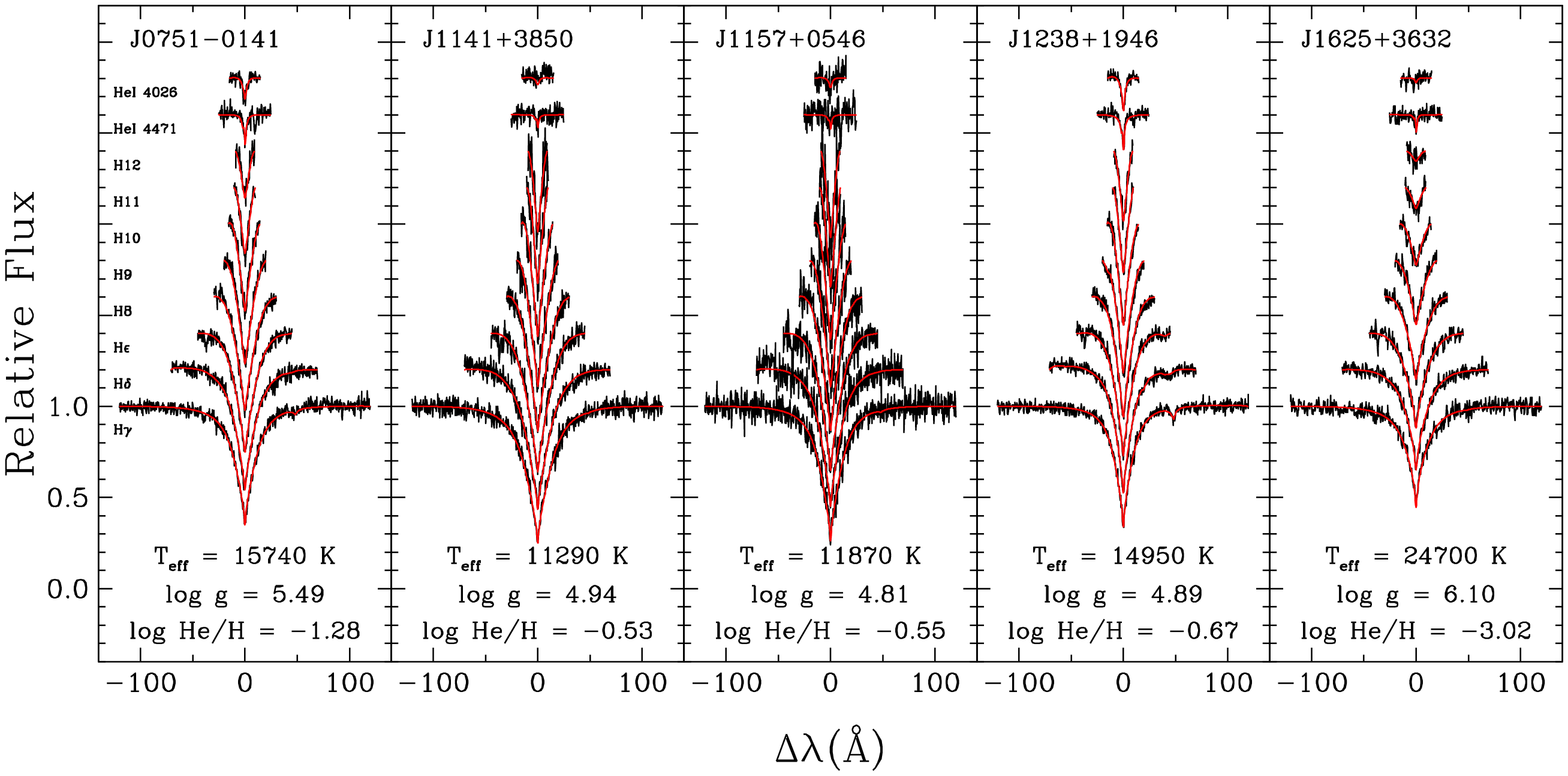}
\figcaption[f02.eps]{Model fits (red) to the hydrogen Balmer line,
  and the \heia\ and \heib\ lines in the observed optical spectra
  (black) of five ELM WDs. The lines range from \hgamma\ (bottom) to
  H12 in addition to \heia\ and \heib\ (top), each offset by a factor
  of 0.2 for clarity. The best-fit values of \Te, \logg, and
  \loghe\ are indicated at the bottom of each panel. \label{fg:DAB}}
\end{figure*}

\section{SPECTROSCOPIC ANALYSIS}

\subsection{Spectroscopic Fits}

Our Balmer line fits use the so-called spectroscopic technique
developed by \citet{bsl92} and described at length in
\citet{gianninas11} and references therein. Briefly, we first
normalize each individual Balmer line to a continuum set to unity, in
both the observed and model spectra. The comparison with the synthetic
spectra, which are convolved with an appropriate Gaussian instrumental
profile (1.0, 1.7, 2.0, 2.3 \AA), is then carried out in terms of
these normalized line shapes only.  Next, we use our grid of model
spectra to determine \Te\ and \logg\ using a minimization technique
which relies on the nonlinear least-squares method of
Levenberg-Marquardt \citep{press86}, which is based on a steepest
descent method. One important difference between our procedure and
that of \citet{gianninas11} is that we fit higher-order Balmer lines
up to and including H12. These higher-order Balmer lines are still
present in low surface gravity ELM WDs and provide additional
constraints on our measurement of \logg. Furthermore, to ensure the
homogeneity of our analysis, we do not fit \hbeta\ for the four
objects whose spectra were obtained at FLWO. We have compared the
\Te\ and \logg\ values from fits with and without \hbeta\ and the
results agree within the uncertainties. Consequently, we fit only the
lines from \hgamma\ to H12 for our entire sample. For the ELM WDs
whose optical spectra also contain lines due to Ca or Mg, we exclude
the affected wavelength ranges from both the normalization and fitting
routines.

\begin{table*}\scriptsize
\caption{ELM WD Physical Parameters}
\begin{center}
\setlength{\tabcolsep}{7.2pt}
\begin{tabular*}{\hsize}{@{\extracolsep{\fill}}lr@{ $\pm$ }@{\extracolsep{0pt}}lr@{ $\pm$ }@{\extracolsep{0pt}}lcr@{ $\pm$ }@{\extracolsep{0pt}}lr@{ $\pm$ }@{\extracolsep{0pt}}lr@{ $\pm$ }@{\extracolsep{0pt}}lr@{ $\pm$ }@{\extracolsep{0pt}}lr@{ $\pm$ }@{\extracolsep{0pt}}l@{}}
\hline
\hline
\noalign{\smallskip}
SDSS & \multicolumn{2}{c}{\Te} & \multicolumn{2}{c}{\logg} & Mass\footnotemark & \multicolumn{2}{c}{Radius} & \multicolumn{2}{c}{$g_{0}$} & \multicolumn{2}{c}{$M_{g}$} & \multicolumn{2}{c}{$d$} & \multicolumn{2}{c}{$\tau_{\rm cool}$} \\
     & \multicolumn{2}{c}{(K)} & \multicolumn{2}{c}{(cm s$^{-2}$)} & (\msun) & \multicolumn{2}{c}{(\rsun)} & \multicolumn{2}{c}{(mag)} & \multicolumn{2}{c}{(mag)} & \multicolumn{2}{c}{(kpc)} & \multicolumn{2}{c}{(Gyr)}         \\
\noalign{\smallskip}
\hline
\noalign{\smallskip}
J0022+0031   & 20460 & 310 & 7.58 & 0.04 & 0.457 & 0.0182 & 0.0013 & 19.284 & 0.033 &  9.84 & 0.19 & 0.775 & 0.068 & 0.215 & 0.129 \\
J0022$-$1014 & 20730 & 340 & 7.28 & 0.05 & 0.375 & 0.0233 & 0.0019 & 19.581 & 0.031 &  9.28 & 0.21 & 1.151 & 0.113 & 0.042 & 0.021 \\
J0056$-$0611 & 12230 & 180 & 6.17 & 0.04 & 0.174 & 0.0565 & 0.0061 & 17.208 & 0.023 &  8.37 & 0.27 & 0.586 & 0.073 & 0.959 & 0.081 \\
J0106$-$1000 & 16970 & 260 & 6.10 & 0.05 & 0.191 & 0.0642 & 0.0068 & 19.595 & 0.023 &  7.45 & 0.26 & 2.690 & 0.323 & 0.497 & 0.168 \\
J0112+1835   & 10020 & 140 & 5.76 & 0.05 & 0.161 & 0.0874 & 0.0108 & 17.110 & 0.016 &  7.89 & 0.31 & 0.697 & 0.100 & 1.661 & 0.169 \\
J0152+0749   & 10840 & 180 & 5.92 & 0.05 & 0.168 & 0.0748 & 0.0084 & 18.033 & 0.009 &  8.03 & 0.29 & 1.001 & 0.134 & 1.384 & 0.130 \\
J0345+1748\footnotemark &  8680 & 120 & 6.83 & 0.04 & 0.220 & 0.0297 & 0.0028 & 16.500 & 0.300 & 10.84 & 0.27 & 0.134 & 0.025 & 1.020 & 0.094 \\
J0651+2844   & 16340 & 260 & 6.81 & 0.05 & 0.252 & 0.0325 & 0.0031 & 19.111 & 0.012 &  9.00 & 0.24 & 1.053 & 0.116 & 0.183 & 0.110 \\
J0730+1703   & 12030 & 220 & 6.30 & 0.05 & 0.183 & 0.0503 & 0.0058 & 20.076 & 0.028 &  8.66 & 0.29 & 1.921 & 0.261 & 1.003 & 0.159 \\
J0745+1949   &  8380 & 130 & 6.21 & 0.07 & 0.164 & 0.0526 & 0.0076 & 16.491 & 0.008 &  9.75 & 0.39 & 0.223 & 0.040 & 4.232 & 0.593 \\
J0751$-$0141 & 15740 & 250 & 5.49 & 0.05 & 0.194 & 0.1315 & 0.0104 & 17.490 & 0.015 &  6.03 & 0.19 & 1.958 & 0.168 & 0.258 & 0.141 \\
J0755+4800   & 19520 & 300 & 7.42 & 0.05 & 0.409 & 0.0207 & 0.0016 & 16.039 & 0.017 &  9.64 & 0.20 & 0.190 & 0.017 & 0.096 & 0.008 \\
J0755+4906   & 13590 & 280 & 6.13 & 0.06 & 0.176 & 0.0597 & 0.0077 & 20.242 & 0.023 &  8.03 & 0.33 & 2.768 & 0.417 & 0.803 & 0.115 \\
J0802$-$0955 & 16980 & 270 & 6.44 & 0.05 & 0.208 & 0.0452 & 0.0047 & 18.885 & 0.012 &  8.21 & 0.26 & 1.366 & 0.161 & 0.356 & 0.160 \\
J0811+0225   & 13540 & 200 & 5.67 & 0.04 & 0.181 & 0.1035 & 0.0111 & 18.669 & 0.024 &  6.84 & 0.26 & 2.321 & 0.284 & 0.485 & 0.071 \\
J0815+2309   & 21430 & 330 & 5.84 & 0.05 & 0.207 & 0.0903 & 0.0092 & 17.805 & 0.015 &  6.27 & 0.25 & 2.025 & 0.234 & 0.380 & 0.197 \\
J0818+3536   & 10360 & 190 & 5.86 & 0.09 & 0.165 & 0.0790 & 0.0128 & 20.756 & 0.026 &  8.03 & 0.40 & 3.512 & 0.659 & 1.613 & 0.183 \\
J0822+2753   &  8980 & 130 & 6.65 & 0.05 & 0.188 & 0.0340 & 0.0037 & 18.314 & 0.013 & 10.42 & 0.30 & 0.380 & 0.053 & 1.129 & 0.169 \\
J0825+1152   & 27180 & 400 & 6.60 & 0.04 & 0.287 & 0.0443 & 0.0038 & 18.774 & 0.018 &  7.34 & 0.22 & 1.938 & 0.198 & 0.083 & 0.089 \\
J0840+1527   & 13670 & 230 & 5.04 & 0.05 & 0.192 & 0.2198 & 0.0241 & 19.319 & 0.027 &  5.19 & 0.27 & 6.711 & 0.846 & 0.181 & 0.030 \\
J0845+1624   & 19620 & 310 & 7.51 & 0.05 & 0.434 & 0.0191 & 0.0015 & 19.817 & 0.020 &  9.81 & 0.20 & 1.003 & 0.093 & 0.130 & 0.056 \\
J0849+0445   & 10290 & 150 & 6.29 & 0.05 & 0.178 & 0.0499 & 0.0060 & 19.292 & 0.020 &  9.05 & 0.30 & 1.119 & 0.156 & 1.617 & 0.150 \\
J0900+0234   &  8490 & 130 & 6.26 & 0.07 & 0.167 & 0.0502 & 0.0071 & 18.142 & 0.016 &  9.79 & 0.38 & 0.467 & 0.082 & 4.188 & 0.660 \\
J0917+4638   & 12240 & 180 & 5.75 & 0.04 & 0.174 & 0.0918 & 0.0099 & 18.764 & 0.019 &  7.31 & 0.27 & 1.956 & 0.242 & 0.792 & 0.095 \\
J0923+3028   & 18500 & 290 & 6.88 & 0.05 & 0.279 & 0.0316 & 0.0028 & 15.709 & 0.019 &  8.83 & 0.23 & 0.238 & 0.025 & 0.079 & 0.075 \\
J1005+0542   & 16590 & 260 & 7.38 & 0.05 & 0.388 & 0.0210 & 0.0017 & 19.763 & 0.023 &  9.93 & 0.20 & 0.927 & 0.087 & 0.166 & 0.025 \\
J1005+3550   & 10060 & 140 & 6.02 & 0.05 & 0.168 & 0.0665 & 0.0078 & 19.004 & 0.610 &  8.48 & 0.30 & 1.273 & 0.402 & 1.861 & 0.115 \\
J1046$-$0153 & 14860 & 230 & 7.37 & 0.05 & 0.375 & 0.0210 & 0.0017 & 18.098 & 0.018 & 10.13 & 0.20 & 0.392 & 0.037 & 0.223 & 0.038 \\
J1053+5200   & 16370 & 240 & 6.54 & 0.04 & 0.213 & 0.0409 & 0.0040 & 18.953 & 0.020 &  8.50 & 0.24 & 1.234 & 0.138 & 0.353 & 0.152 \\
J1056+6536   & 21010 & 360 & 7.10 & 0.05 & 0.338 & 0.0272 & 0.0024 & 19.784 & 0.023 &  8.92 & 0.22 & 1.489 & 0.153 & 0.009 & 0.036 \\
J1104+0918   & 16700 & 260 & 7.61 & 0.05 & 0.454 & 0.0174 & 0.0013 & 16.659 & 0.016 & 10.32 & 0.19 & 0.186 & 0.017 & 0.267 & 0.178 \\
J1112+1117   &  9560 & 140 & 6.30 & 0.06 & 0.177 & 0.0490 & 0.0061 & 16.307 & 0.016 &  9.33 & 0.33 & 0.248 & 0.038 & 2.428 & 0.347 \\
J1141+3850   & 11290 & 210 & 4.94 & 0.10 & 0.177 & 0.2358 & 0.0392 & 19.058 & 0.017 &  5.42 & 0.39 & 5.330 & 0.967 & 0.233 & 0.074 \\
J1151+5858   & 15430 & 300 & 6.10 & 0.06 & 0.183 & 0.0632 & 0.0077 & 20.150 & 0.025 &  7.66 & 0.30 & 3.150 & 0.441 & 0.649 & 0.149 \\
J1157+0546   & 11870 & 260 & 4.81 & 0.14 & 0.186 & 0.2798 & 0.0575 & 19.818 & 0.024 &  4.94 & 0.49 & 9.432 & 2.140 & 0.152 & 0.056 \\
J1233+1602   & 11700 & 240 & 5.59 & 0.07 & 0.169 & 0.1092 & 0.0148 & 19.911 & 0.017 &  7.03 & 0.34 & 3.777 & 0.599 & 0.781 & 0.108 \\
J1234$-$0228 & 17800 & 260 & 6.61 & 0.04 & 0.229 & 0.0391 & 0.0037 & 17.855 & 0.016 &  8.43 & 0.23 & 0.766 & 0.082 & 0.285 & 0.142 \\
J1238+1946   & 14950 & 240 & 4.89 & 0.05 & 0.210 & 0.2716 & 0.0226 & 17.291 & 0.023 &  4.55 & 0.19 & 3.529 & 0.319 & 0.104 & 0.029 \\
J1422+4352   & 12770 & 250 & 6.11 & 0.06 & 0.174 & 0.0606 & 0.0077 & 19.822 & 0.023 &  8.12 & 0.32 & 2.187 & 0.323 & 0.875 & 0.102 \\
J1436+5010   & 17370 & 250 & 6.66 & 0.04 & 0.233 & 0.0375 & 0.0035 & 18.236 & 0.015 &  8.58 & 0.23 & 0.855 & 0.091 & 0.247 & 0.146 \\
J1439+1002   & 14520 & 220 & 6.36 & 0.05 & 0.185 & 0.0471 & 0.0050 & 17.938 & 0.012 &  8.42 & 0.26 & 0.803 & 0.098 & 0.544 & 0.161 \\
J1443+1509   &  9170 & 130 & 6.71 & 0.06 & 0.200 & 0.0328 & 0.0038 & 18.650 & 0.016 & 10.41 & 0.32 & 0.445 & 0.065 & 0.977 & 0.125 \\
J1448+1342   & 12300 & 360 & 6.99 & 0.06 & 0.269 & 0.0274 & 0.0028 & 19.286 & 0.023 &  9.94 & 0.29 & 0.738 & 0.098 & 0.336 & 0.089 \\
J1512+2615   & 11380 & 180 & 6.93 & 0.06 & 0.251 & 0.0284 & 0.0031 & 19.474 & 0.019 & 10.03 & 0.27 & 0.774 & 0.097 & 0.389 & 0.074 \\
J1518+0658   &  9940 & 140 & 6.82 & 0.04 & 0.224 & 0.0306 & 0.0030 & 17.581 & 0.017 & 10.24 & 0.27 & 0.293 & 0.036 & 0.714 & 0.053 \\
J1538+0252   & 10260 & 150 & 5.98 & 0.06 & 0.168 & 0.0694 & 0.0093 & 18.721 & 0.014 &  8.33 & 0.33 & 1.195 & 0.183 & 1.726 & 0.106 \\
J1557+2823   & 12560 & 190 & 7.76 & 0.05 & 0.461 & 0.0147 & 0.0011 & 17.712 & 0.029 & 11.26 & 0.20 & 0.195 & 0.018 & 0.747 & 0.480 \\
J1614+1912   &  8870 & 160 & 6.62 & 0.13 & 0.186 & 0.0348 & 0.0072 & 16.395 & 0.019 & 10.42 & 0.55 & 0.157 & 0.040 & 1.435 & 1.177 \\
J1625+3632   & 24700 & 400 & 6.10 & 0.05 & 0.210 & 0.0678 & 0.0053 & 19.370 & 0.016 &  6.62 & 0.18 & 3.547 & 0.303 & 0.455 & 0.226 \\
J1630+2712   & 10390 & 170 & 6.03 & 0.08 & 0.171 & 0.0663 & 0.0099 & 20.146 & 0.017 &  8.40 & 0.37 & 2.230 & 0.382 & 1.669 & 0.105 \\
J1630+4233   & 16070 & 250 & 7.07 & 0.05 & 0.307 & 0.0266 & 0.0023 & 19.071 & 0.017 &  9.47 & 0.22 & 0.832 & 0.085 & 0.128 & 0.020 \\
J1741+6526   & 10540 & 170 & 6.00 & 0.06 & 0.170 & 0.0678 & 0.0088 & 18.370 & 0.021 &  8.32 & 0.32 & 1.025 & 0.154 & 1.578 & 0.114 \\
J1840+6423   &  9380 & 130 & 6.55 & 0.05 & 0.183 & 0.0377 & 0.0042 & 18.963 & 0.014 & 10.00 & 0.31 & 0.621 & 0.089 & 1.561 & 0.361 \\
J2103$-$0027 & 10130 & 150 & 5.78 & 0.05 & 0.162 & 0.0861 & 0.0103 & 18.488 & 0.014 &  7.90 & 0.30 & 1.313 & 0.182 & 1.633 & 0.156 \\
J2119$-$0018 &  9980 & 150 & 5.71 & 0.08 & 0.160 & 0.0923 & 0.0140 & 20.171 & 0.022 &  7.79 & 0.38 & 2.997 & 0.522 & 1.588 & 0.187 \\
J2132+0754   & 13780 & 200 & 6.02 & 0.04 & 0.177 & 0.0681 & 0.0073 & 18.105 & 0.019 &  7.72 & 0.26 & 1.196 & 0.146 & 0.718 & 0.140 \\
J2228+3623   &  7990 & 120 & 6.07 & 0.08 & 0.153 & 0.0599 & 0.0094 & 16.964 & 0.011 &  9.68 & 0.42 & 0.286 & 0.056 & 5.513 & 2.911 \\
J2236+2232\footnotemark & 11360 & 170 & 6.53 & 0.04 & 0.182 & 0.0384 & 0.0040 & 17.163 & 0.019 &  9.38 & 0.26 & 0.359 & 0.044 & 1.106 & 0.174 \\
J2252$-$0056 & 21690 & 310 & 7.15 & 0.04 & 0.352 & 0.0262 & 0.0020 & 18.591 & 0.026 &  8.94 & 0.20 & 0.853 & 0.078 & 0.013 & 0.029 \\
J2338$-$2052 & 16620 & 280 & 6.85 & 0.05 & 0.263 & 0.0318 & 0.0030 & 19.674 & 0.035 &  9.02 & 0.24 & 1.354 & 0.152 & 0.146 & 0.065 \\
J2345$-$0102 & 34270 & 500 & 7.42 & 0.05 & 0.466 & 0.0219 & 0.0010 & 19.539 & 0.020 &  8.36 & 0.13 & 1.719 & 0.103 & 0.024 & 0.001 \\
\noalign{\smallskip}
\hline
\noalign{\smallskip}
\multicolumn{16}{@{}l}{$^{\rm a}$We adopt an uncertainty of 0.020~\msun\ for all estimates of the primary mass.}\\
\multicolumn{16}{@{}l}{$^{\rm b}$\nltt; since this WD is outside the SDSS footprint, we adopt the $V$ magnitude from \citet{kawka09} instead of $g_{0}$} \\ 
\multicolumn{16}{@{}l}{$^{\rm c}$\lp} \\
\end{tabular*}
\label{tab:par}
\end{center}
\end{table*}

\begin{figure*}[!ht]
\includegraphics[scale=0.775,angle=-90,bb=62 46 576 784]{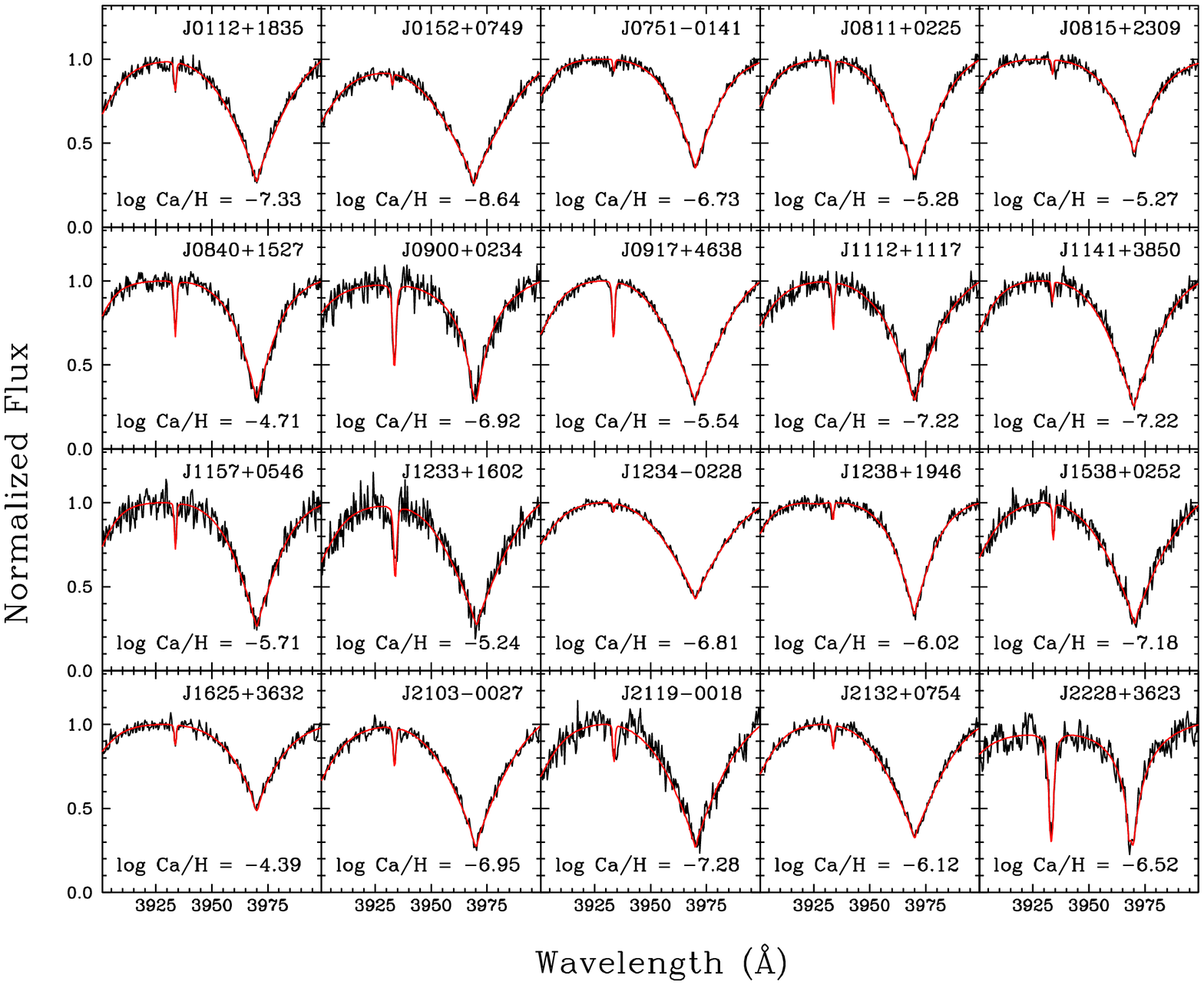}
\figcaption[f03.ps]{Fits of the Ca~{\sc II} H and K lines. Note
  that the H line (\caiih) is blended with \hepsilon. The measured
  Ca abundance for each ELM WD is indicated in the individual
  panels.  \label{fg:Ca}}
\includegraphics[scale=0.575,angle=-90,bb=32 -76 446 634]{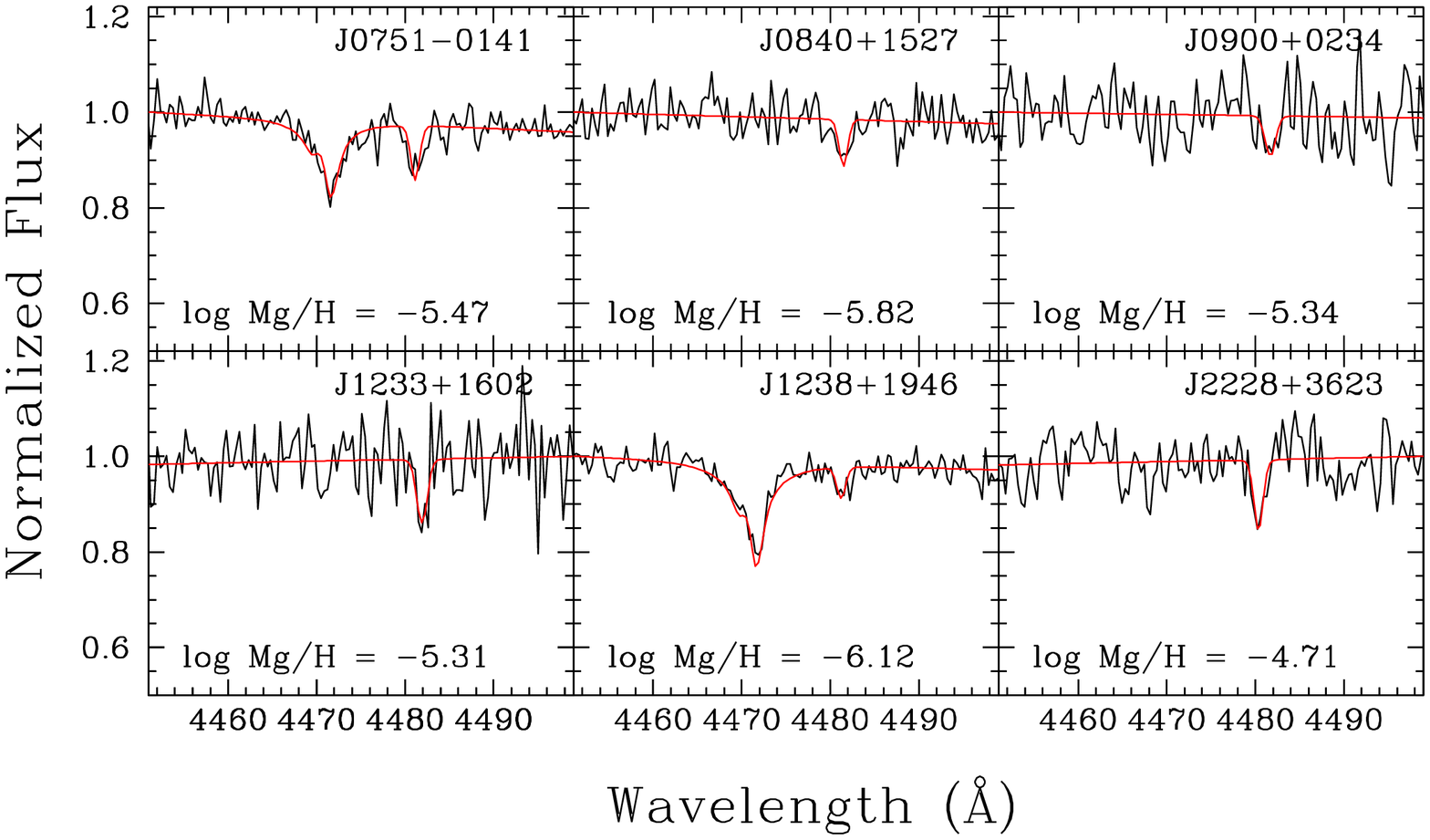}
\figcaption[f04.eps]{Same as Figure \ref{fg:Ca} but for the 
  \mgii\ doublet. In the panels for J0751, and J1238 the
  \heii\ line can also be observed. \label{fg:Mg}}
\end{figure*}

\begin{figure}[!h]
\centering
\includegraphics[scale=0.425,bb=20 117 592 659]{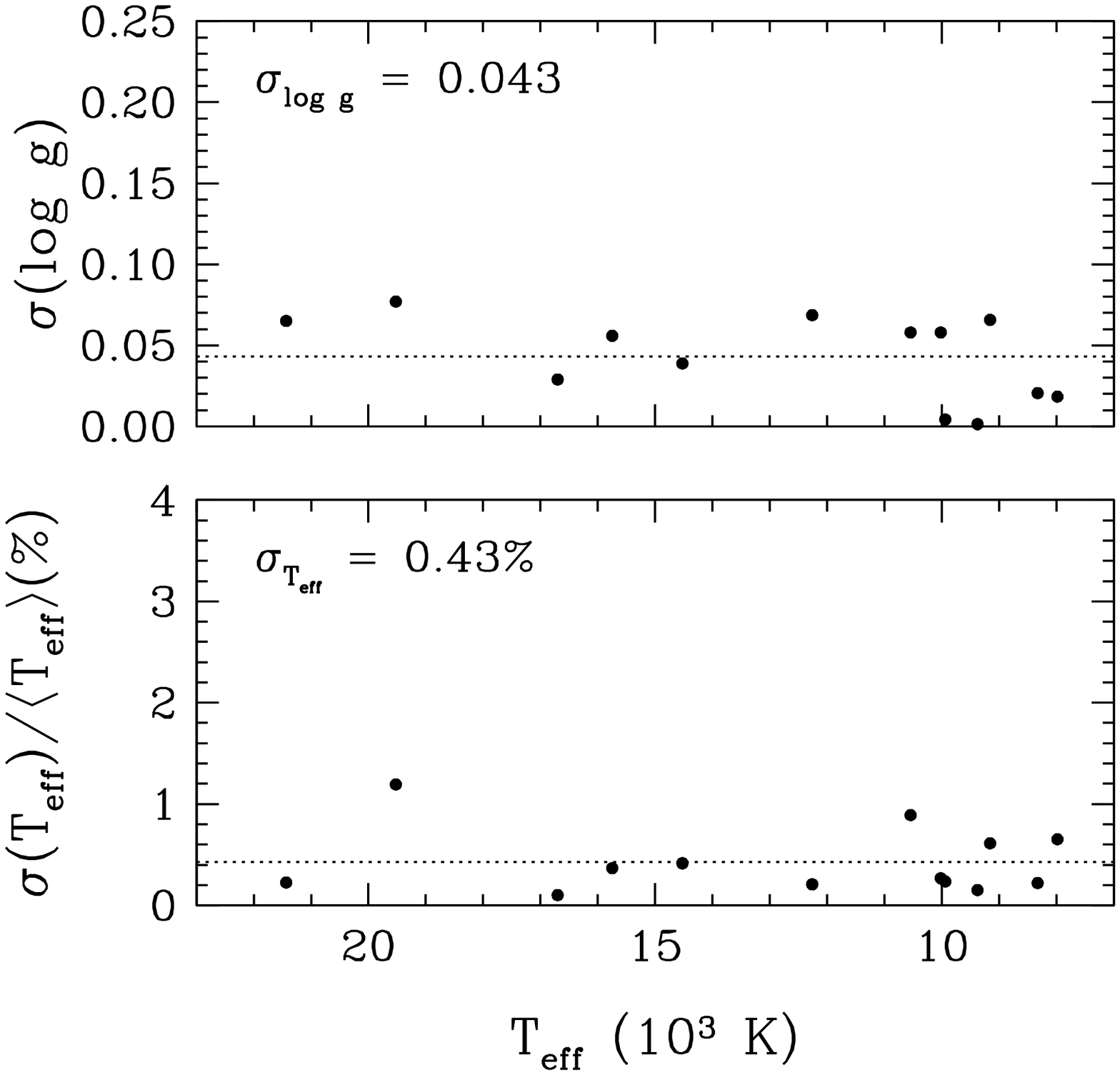}
\figcaption[f05.eps]{Distribution of standard deviations in \Te\ and
  \logg\ for individual ELM WDs with multiple measurements as a
  function of \Te. Standard deviations in \Te\ are expressed in
  percentage with respect to the average temperature of the star.  The
  dotted lines represent the average standard deviations as indicated
  in the top left of each panel.  \label{fg:sigma}}
\end{figure}

\begin{figure}[!h]
\centering
\includegraphics[scale=0.355,angle=-90,bb=122 51 546 784]{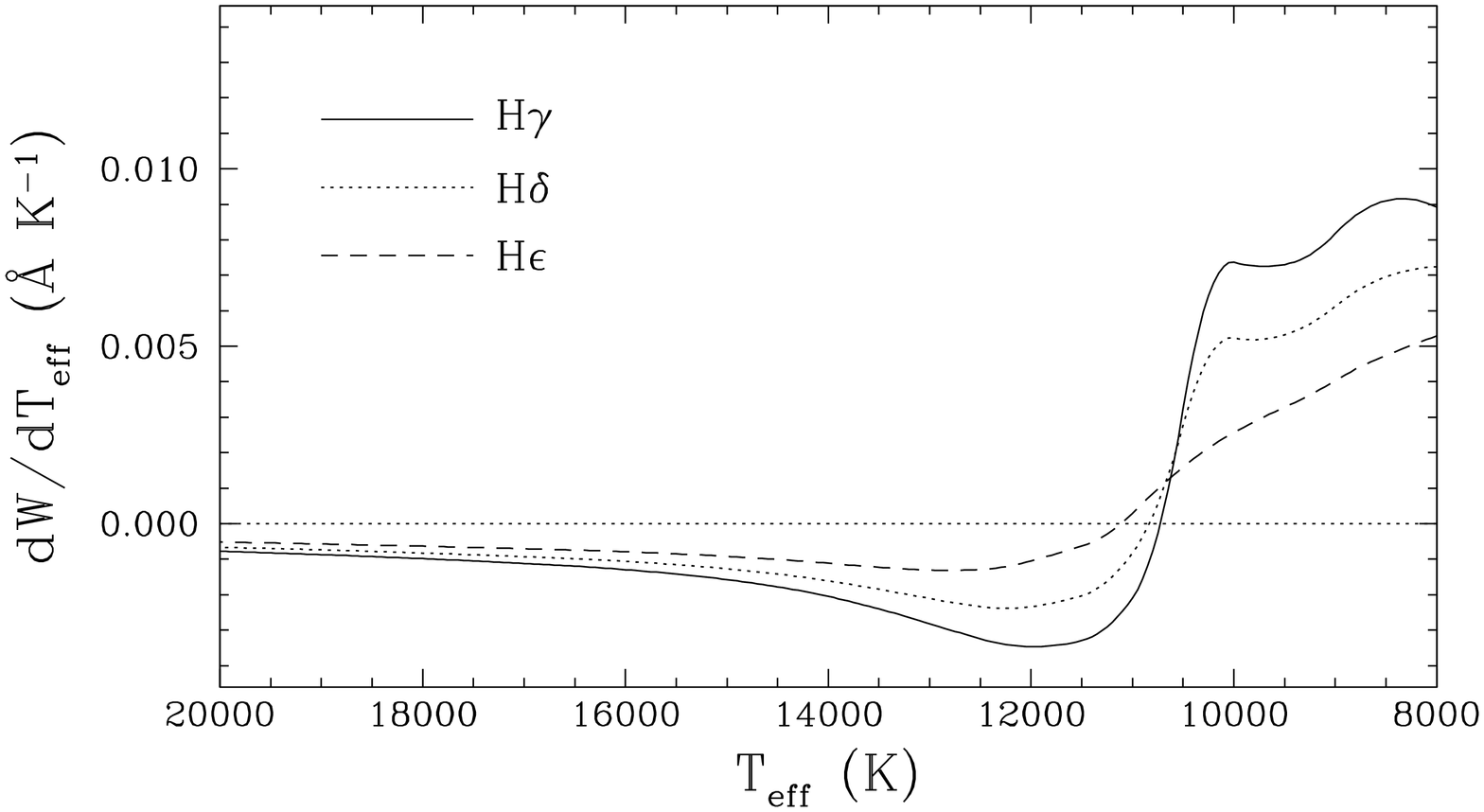}
\figcaption[f06.eps]{Derivative of the equivalent width of \hgamma,
  \hdelta, and \hepsilon\ with respect to \Te\ as a function of the
  effective temperature for models of ELM WDs. The surface gravity is
  held fixed here at a value of \logg\ = 6.0. \label{fg:EW}}
\end{figure}

\begin{table*}\scriptsize
\caption{ELM WD Binary Parameters}
\begin{center}
\setlength{\tabcolsep}{8.15pt}
\begin{tabular*}{\hsize}{@{\extracolsep{\fill}}lr@{ $\pm$ }@{\extracolsep{0pt}}lr@{ $\pm$ }@{\extracolsep{0pt}}lr@{ $\pm$ }@{\extracolsep{0pt}}lr@{ $\pm$ }@{\extracolsep{0pt}}lr@{ $\pm$ }@{\extracolsep{0pt}}lcr@{ $\pm$ }@{\extracolsep{0pt}}lc@{}}
\hline
\hline
\noalign{\smallskip}
SDSS & \multicolumn{2}{c}{$P$} & \multicolumn{2}{c}{$K$} & \multicolumn{2}{c}{Mass Function} & \multicolumn{2}{c}{$M_{2}$} & \multicolumn{2}{c}{$M_{2, i=60^{\circ}}$} & $\tau_{\rm merge}$ & \multicolumn{2}{c}{$a$} & $\log h$ \\
     & \multicolumn{2}{c}{(days)}       & \multicolumn{2}{c}{(\kms)} & \multicolumn{2}{c}{(\msun)} & \multicolumn{2}{c}{(\msun)} & \multicolumn{2}{c}{(\msun)} & (Gyr) & \multicolumn{2}{c}{(\rsun)} & \\
\noalign{\smallskip}
\hline
\noalign{\smallskip}
J0022+0031   & 0.49135 & 0.02540 &  80.8 & 1.3 & 0.027 & 0.003 & $\geqslant$0.23 & 0.02 & 0.28 & 0.02 & \ldots            & 2.32 & 0.12 & $-$22.79 \\
J0022$-$1014 & 0.07989 & 0.00300 & 145.6 & 5.6 & 0.026 & 0.004 & $\geqslant$0.20 & 0.02 & 0.25 & 0.02 & $\leqslant$ 0.616 & 0.65 & 0.03 & $-$22.55 \\
J0056$-$0611 & 0.04338 & 0.00002 & 376.9 & 2.4 & 0.241 & 0.005 & $\geqslant$0.46 & 0.02 & 0.61 & 0.03 & $\leqslant$ 0.120 & 0.45 & 0.01 & $-$22.06 \\
J0106$-$1000 & 0.02715 & 0.00002 & 395.2 & 3.6 & 0.174 & 0.005 & $\geqslant$0.39 & 0.02 & 0.51 & 0.03 & $\leqslant$ 0.036 & 0.32 & 0.01 & $-$22.61 \\
J0112+1835   & 0.14698 & 0.00003 & 295.3 & 2.0 & 0.392 & 0.008 & $\geqslant$0.62 & 0.03 & 0.85 & 0.04 & $\leqslant$ 2.650 & 1.08 & 0.02 & $-$22.41 \\
J0152+0749   & 0.32288 & 0.00014 & 217.0 & 2.0 & 0.342 & 0.010 & $\geqslant$0.57 & 0.03 & 0.78 & 0.04 & \ldots            & 1.79 & 0.04 & $-$22.81 \\
J0345+1748   & 0.23503 & 0.00013 & 273.4 & 0.5 & 0.498 & 0.003 & $\geqslant$0.81 & 0.03 & 0.72 & 0.01$^{\rm a}$ & $\leqslant$ 5.742 & 1.62 & 0.02 & $-$21.97 \\
J0651+2844   & 0.00886 & 0.00001 & 616.9 & 5.0 & 0.215 & 0.005 & $\geqslant$0.49 & 0.02 & 0.50 & 0.04$^{\rm a}$ & $\leqslant$ 0.001 & 0.16 & 0.01 & $-$21.97 \\
J0730+1703   & 0.69770 & 0.05427 & 122.8 & 4.3 & 0.134 & 0.025 & $\geqslant$0.33 & 0.05 & 0.42 & 0.07 & \ldots            & 2.64 & 0.26 & $-$23.48 \\
J0745+1949   & 0.11240 & 0.00833 & 108.7 & 2.9 & 0.015 & 0.002 & $\geqslant$0.10 & 0.01 & 0.12 & 0.02 & $\leqslant$ 5.448 & 0.63 & 0.06 & $-$22.49 \\
J0751$-$0141 & 0.08001 & 0.00279 & 432.6 & 2.3 & 0.671 & 0.034 & $\geqslant$0.97 & 0.06 & 0.97 & 0.04${\rm ^a}$ & $\leqslant$ 0.320 & 0.82 & 0.04 & $-$22.77 \\
J0755+4800   & 0.54627 & 0.00522 & 194.5 & 5.5 & 0.416 & 0.039 & $\geqslant$0.89 & 0.07 & 1.17 & 0.09 & \ldots            & 3.07 & 0.09 & $-$21.75 \\
J0755+4906   & 0.06302 & 0.00213 & 438.0 & 5.0 & 0.549 & 0.037 & $\geqslant$0.81 & 0.06 & 1.13 & 0.09 & $\leqslant$ 0.210 & 0.66 & 0.03 & $-$22.64 \\
J0802$-$0955 & 0.54687 & 0.00455 & 176.5 & 4.5 & 0.312 & 0.026 & $\geqslant$0.58 & 0.05 & 0.77 & 0.07 & \ldots            & 2.60 & 0.09 & $-$23.01 \\
J0811+0225   & 0.82194 & 0.00049 & 220.7 & 2.5 & 0.915 & 0.032 & $\geqslant$1.21 & 0.06 & 1.72 & 0.08 & \ldots            & 4.12 & 0.08 & $-$23.17 \\
J0815+2309   & 1.07357 & 0.00018 & 131.7 & 2.6 & 0.254 & 0.015 & $\geqslant$0.50 & 0.04 & 0.67 & 0.05 & \ldots            & 3.94 & 0.11 & $-$23.43 \\
J0818+3536   & 0.18315 & 0.02110 & 170.0 & 5.0 & 0.093 & 0.019 & $\geqslant$0.25 & 0.04 & 0.33 & 0.06 & $\leqslant$ 9.269 & 1.02 & 0.13 & $-$23.48 \\
J0822+2753   & 0.24400 & 0.00020 & 271.1 & 9.0 & 0.504 & 0.051 & $\geqslant$0.78 & 0.08 & 1.07 & 0.11 & $\leqslant$ 7.529 & 1.62 & 0.06 & $-$22.16 \\
J0825+1152   & 0.05819 & 0.00001 & 319.4 & 2.7 & 0.196 & 0.005 & $\geqslant$0.49 & 0.02 & 0.64 & 0.03 & $\leqslant$ 0.158 & 0.58 & 0.01 & $-$22.45 \\
J0840+1527   & 0.52155 & 0.00474 &  84.8 & 3.1 & 0.033 & 0.004 & $\geqslant$0.16 & 0.02 & 0.20 & 0.02 & \ldots            & 1.92 & 0.08 & $-$24.19 \\
J0845+1624   & 0.75599 & 0.02164 &  62.2 & 5.4 & 0.019 & 0.005 & $\geqslant$0.19 & 0.03 & 0.23 & 0.04 & \ldots            & 2.99 & 0.13 & $-$23.12 \\
J0849+0445   & 0.07870 & 0.00010 & 366.9 & 4.7 & 0.403 & 0.016 & $\geqslant$0.65 & 0.04 & 0.89 & 0.05 & $\leqslant$ 0.441 & 0.73 & 0.02 & $-$22.38 \\
J0900+0234   & \multicolumn{2}{c}{\ldots} & \multicolumn{2}{c}{$\leqslant$24.0} & \multicolumn{2}{c}{\ldots}& \multicolumn{2}{c}{\ldots} & \multicolumn{2}{c}{\ldots} & \ldots & \multicolumn{2}{c}{\ldots} & \ldots \\
J0917+4638   & 0.31642 & 0.00002 & 148.8 & 2.0 & 0.108 & 0.004 & $\geqslant$0.28 & 0.02 & 0.36 & 0.02 & \ldots            & 1.50 & 0.04 & $-$23.33 \\
J0923+3028   & 0.04495 & 0.00049 & 296.0 & 3.0 & 0.121 & 0.005 & $\geqslant$0.37 & 0.02 & 0.47 & 0.03 & $\leqslant$ 0.102 & 0.46 & 0.01 & $-$21.58 \\
J1005+0542   & 0.30560 & 0.00007 & 208.9 & 6.8 & 0.289 & 0.028 & $\geqslant$0.70 & 0.05 & 0.91 & 0.07 & $\leqslant$ 7.696 & 1.96 & 0.04 & $-$22.38 \\
J1005+3550   & 0.17652 & 0.00011 & 143.0 & 2.3 & 0.053 & 0.003 & $\geqslant$0.19 & 0.02 & 0.24 & 0.02 & $\leqslant$10.448 & 0.94 & 0.03 & $-$23.13 \\
J1046$-$0153 & 0.39539 & 0.10836 &  80.8 & 6.6 & 0.022 & 0.011 & $\geqslant$0.19 & 0.05 & 0.23 & 0.06 & \ldots            & 1.87 & 0.42 & $-$22.58 \\
J1053+5200   & 0.04256 & 0.00002 & 264.0 & 2.0 & 0.081 & 0.002 & $\geqslant$0.26 & 0.01 & 0.33 & 0.02 & $\leqslant$ 0.147 & 0.40 & 0.01 & $-$22.50 \\
J1056+6536   & 0.04351 & 0.00103 & 267.5 & 7.4 & 0.086 & 0.009 & $\geqslant$0.34 & 0.03 & 0.43 & 0.04 & $\leqslant$ 0.085 & 0.46 & 0.02 & $-$22.33 \\
J1104+0918   & 0.55319 & 0.00502 & 142.1 & 6.0 & 0.164 & 0.022 & $\geqslant$0.55 & 0.05 & 0.69 & 0.07 & \ldots            & 2.84 & 0.08 & $-$21.88 \\
J1112+1117   & 0.17248 & 0.00001 & 116.2 & 2.8 & 0.028 & 0.002 & $\geqslant$0.14 & 0.01 & 0.17 & 0.02 & $\leqslant$12.019 & 0.89 & 0.03 & $-$22.51 \\
J1141+3850   & 0.25958 & 0.00005 & 265.8 & 3.5 & 0.505 & 0.020 & $\geqslant$0.76 & 0.05 & 1.06 & 0.06 & $\leqslant$ 9.518 & 1.68 & 0.04 & $-$23.36 \\
J1151+5858   & 0.66902 & 0.00070 & 175.7 & 5.9 & 0.376 & 0.038 & $\geqslant$0.63 & 0.06 & 0.85 & 0.09 & \ldots            & 3.00 & 0.11 & $-$23.45 \\
J1157+0546   & 0.56500 & 0.01925 & 158.3 & 4.9 & 0.232 & 0.030 & $\geqslant$0.46 & 0.05 & 0.61 & 0.07 & \ldots            & 2.49 & 0.15 & $-$23.98 \\
J1233+1602   & 0.15090 & 0.00009 & 336.0 & 4.0 & 0.593 & 0.022 & $\geqslant$0.85 & 0.05 & 1.19 & 0.06 & $\leqslant$ 2.161 & 1.20 & 0.03 & $-$23.03 \\
J1234$-$0228 & 0.09143 & 0.00400 &  94.0 & 2.3 & 0.008 & 0.001 & $\geqslant$0.09 & 0.01 & 0.11 & 0.01 & $\leqslant$ 2.606 & 0.59 & 0.03 & $-$22.89 \\
J1238+1946   & 0.22275 & 0.00009 & 258.6 & 2.5 & 0.399 & 0.012 & $\geqslant$0.68 & 0.03 & 0.92 & 0.04 & $\leqslant$ 5.871 & 1.49 & 0.03 & $-$23.11 \\
J1422+4352   & 0.37930 & 0.01123 & 176.0 & 6.0 & 0.214 & 0.028 & $\geqslant$0.42 & 0.05 & 0.56 & 0.07 & \ldots            & 1.86 & 0.11 & $-$23.29 \\
J1436+5010   & 0.04580 & 0.00010 & 347.4 & 8.9 & 0.199 & 0.016 & $\geqslant$0.45 & 0.04 & 0.59 & 0.05 & $\leqslant$ 0.107 & 0.47 & 0.01 & $-$22.13 \\
J1439+1002   & 0.43741 & 0.00169 & 174.0 & 2.0 & 0.239 & 0.009 & $\geqslant$0.47 & 0.03 & 0.62 & 0.04 & \ldots            & 2.10 & 0.06 & $-$22.84 \\
J1443+1509   & 0.19053 & 0.02402 & 306.7 & 3.0 & 0.569 & 0.089 & $\geqslant$0.86 & 0.12 & 1.19 & 0.17 & $\leqslant$ 3.403 & 1.42 & 0.18 & $-$22.10 \\
J1448+1342   & \multicolumn{2}{c}{\ldots} & \multicolumn{2}{c}{$\leqslant$35.0} & \multicolumn{2}{c}{\ldots} & \multicolumn{2}{c}{\ldots} & \multicolumn{2}{c}{\ldots} & \ldots & \multicolumn{2}{c}{\ldots} & \ldots \\
J1512+2615   & 0.59999 & 0.02348 & 115.0 & 4.0 & 0.095 & 0.014 & $\geqslant$0.31 & 0.03 & 0.39 & 0.04 & \ldots            & 2.47 & 0.14 & $-$22.95 \\
J1518+0658   & 0.60935 & 0.00004 & 172.0 & 2.0 & 0.321 & 0.011 & $\geqslant$0.60 & 0.03 & 0.81 & 0.04 & \ldots            & 2.84 & 0.06 & $-$22.33 \\
J1538+0252   & 0.41915 & 0.00295 & 227.6 & 4.9 & 0.512 & 0.037 & $\geqslant$0.76 & 0.06 & 1.06 & 0.09 & \ldots            & 2.30 & 0.08 & $-$22.87 \\
J1557+2823   & 0.40741 & 0.00294 & 131.2 & 4.2 & 0.095 & 0.010 & $\geqslant$0.42 & 0.03 & 0.52 & 0.04 & \ldots            & 2.22 & 0.05 & $-$21.91 \\
J1614+1912   & \multicolumn{2}{c}{\ldots} & \multicolumn{2}{c}{$\leqslant$56.0} & \multicolumn{2}{c}{\ldots} & \multicolumn{2}{c}{\ldots} & \multicolumn{2}{c}{\ldots} & \ldots & \multicolumn{2}{c}{\ldots} & \ldots \\
J1625+3632   & 0.23238 & 0.03960 &  58.4 & 2.7 & 0.005 & 0.001 & $\geqslant$0.07 & 0.01 & 0.09 & 0.02 & \ldots            & 1.04 & 0.16 & $-$23.95 \\
J1630+2712   & 0.27646 & 0.00002 & 218.0 & 5.0 & 0.297 & 0.020 & $\geqslant$0.52 & 0.04 & 0.70 & 0.06 & \ldots            & 1.58 & 0.05 & $-$23.14 \\
J1630+4233   & 0.02766 & 0.00004 & 295.9 & 4.9 & 0.074 & 0.004 & $\geqslant$0.30 & 0.02 & 0.38 & 0.02 & $\leqslant$ 0.031 & 0.33 & 0.01 & $-$22.03 \\
J1741+6526   & 0.06111 & 0.00001 & 508.0 & 4.0 & 0.830 & 0.020 & $\geqslant$1.10 & 0.05 & 1.57 & 0.06 & $\leqslant$ 0.160 & 0.71 & 0.01 & $-$22.12 \\
J1840+6423   & 0.19130 & 0.00005 & 272.0 & 2.0 & 0.399 & 0.009 & $\geqslant$0.65 & 0.03 & 0.89 & 0.04 & $\leqslant$ 4.579 & 1.32 & 0.03 & $-$22.37 \\
J2103$-$0027 & 0.20308 & 0.00023 & 281.0 & 3.2 & 0.467 & 0.016 & $\geqslant$0.71 & 0.04 & 0.98 & 0.05 & $\leqslant$ 5.683 & 1.39 & 0.03 & $-$22.74 \\
J2119$-$0018 & 0.08677 & 0.00004 & 383.0 & 4.0 & 0.505 & 0.016 & $\geqslant$0.74 & 0.04 & 1.04 & 0.05 & $\leqslant$ 0.574 & 0.80 & 0.02 & $-$22.84 \\
J2132+0754   & 0.25056 & 0.00002 & 297.3 & 3.0 & 0.682 & 0.021 & $\geqslant$0.96 & 0.05 & 1.34 & 0.06 & $\leqslant$ 7.337 & 1.74 & 0.03 & $-$22.62 \\
J2228+3623   & \multicolumn{2}{c}{\ldots} & \multicolumn{2}{c}{$\leqslant$28.0} & \multicolumn{2}{c}{\ldots} & \multicolumn{2}{c}{\ldots} & \multicolumn{2}{c}{\ldots} & \ldots & \multicolumn{2}{c}{\ldots} & \ldots \\
J2236+2232   & 1.01016 & 0.00005 & 119.9 & 2.0 & 0.180 & 0.009 & $\geqslant$0.39 & 0.03 & 0.51 & 0.04 & \ldots            & 3.52 & 0.10 & $-$22.80 \\
J2252$-$0056 & \multicolumn{2}{c}{\ldots} & \multicolumn{2}{c}{$\leqslant$25.0} & \multicolumn{2}{c}{\ldots} & \multicolumn{2}{c}{\ldots} & \multicolumn{2}{c}{\ldots} & \ldots & \multicolumn{2}{c}{\ldots} & \ldots \\
J2338$-$2052 & 0.07644 & 0.00712 & 133.4 & 7.5 & 0.019 & 0.005 & $\geqslant$0.15 & 0.02 & 0.18 & 0.03 & $\leqslant$ 0.972 & 0.56 & 0.05 & $-$22.86 \\
J2345$-$0102 & \multicolumn{2}{c}{\ldots} & \multicolumn{2}{c}{$\leqslant$43.0} & \multicolumn{2}{c}{\ldots} & \multicolumn{2}{c}{\ldots} & \multicolumn{2}{c}{\ldots} & \ldots & \multicolumn{2}{c}{\ldots} & \ldots \\
\noalign{\smallskip}
\hline
\noalign{\smallskip}
\multicolumn{15}{@{}l}{$^{\rm a}$Eclipsing systems where we adopt $M_{2}$ and $i$ as determined from the eclipse modeling (see Section 4.2) instead of assuming $i=60^{\circ}$.}\\
\end{tabular*}
\label{tab:bin}
\end{center}
\end{table*}

\begin{figure}[!ht]
\includegraphics[scale=0.65,bb=90 37 592 749]{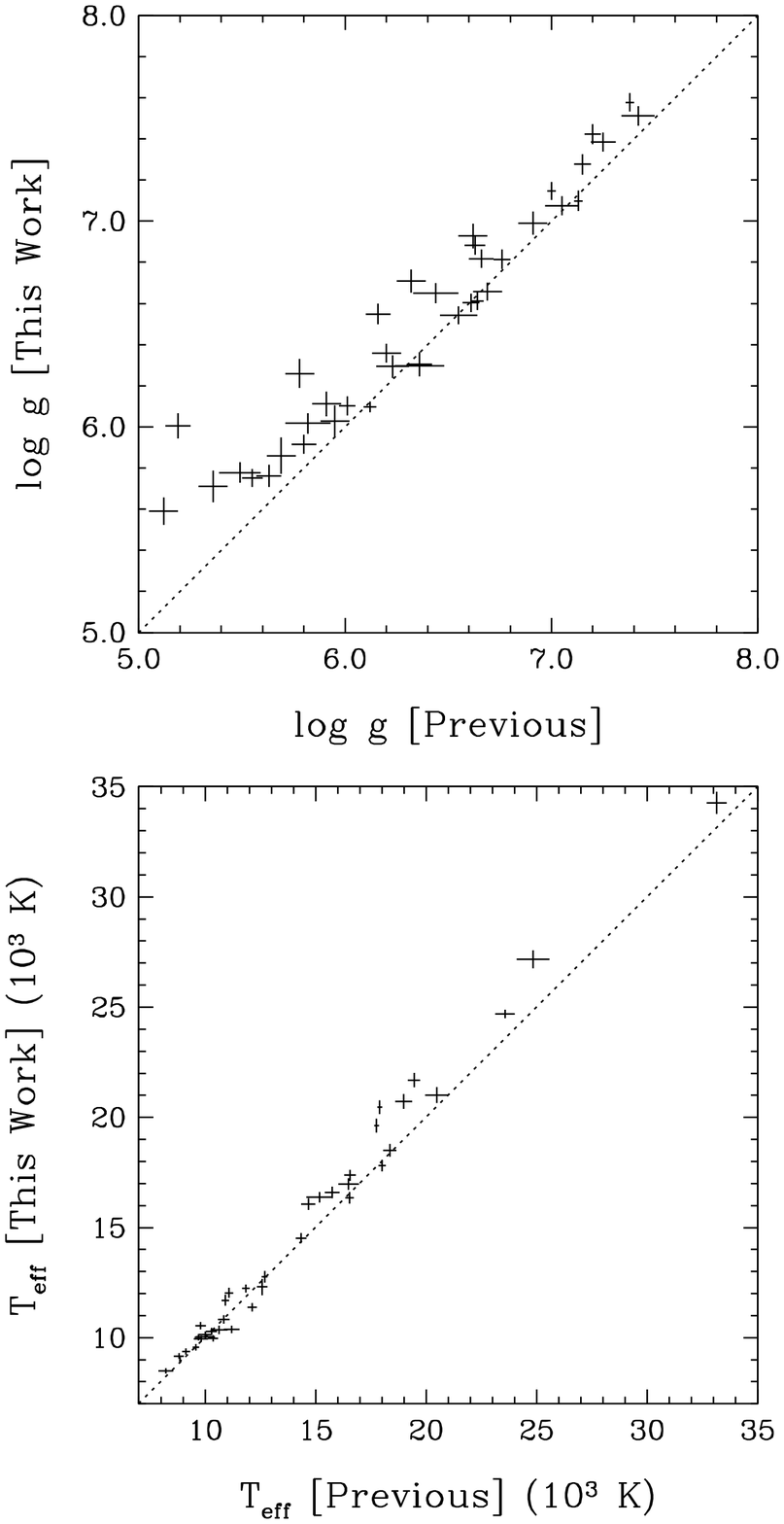}
\figcaption[f07.eps]{Comparison of the atmospheric parameters
  \Te\ (bottom) and \logg\ (top) presented in this work to those
  published in previous analyses \citep[see Table 5 of][and references
    therein]{brown_ELM5}. In both panels, the dashed line represents
  the 1:1 correlation.  \label{fg:comp}}
\end{figure}

\begin{table}[!t]\small
\caption{Measured Atmospheric Abundances}
\begin{center}
\begin{tabular}{@{\extracolsep{\fill}}lccc@{}}
\hline
\hline
\noalign{\smallskip}
SDSS & \loghe & \logmg & \logca \\
\noalign{\smallskip}
\hline
\noalign{\smallskip}
J0112+1835   & \ldots  & \ldots  & $-$7.33 \\
J0152+0749   & \ldots  & \ldots  & $-$8.64 \\
J0745+1949   & \ldots  & $-$3.90 & $-$5.80 \\
J0751$-$0141 & $-$1.28 & $-$5.47 & $-$6.73 \\
J0811+0225   & \ldots  & \ldots  & $-$5.28 \\
J0815+2309   & \ldots  & \ldots  & $-$5.27 \\
J0840+1527   & \ldots  & $-$5.82 & $-$4.71 \\
J0900+0234   & \ldots  & $-$5.34 & $-$6.92 \\
J0917+4638   & \ldots  & \ldots  & $-$5.54 \\
J1112+1117   & \ldots  & \ldots  & $-$7.22 \\
J1141+3850   & $-$0.53 & \ldots  & $-$7.22 \\
J1157+0546   & $-$0.55 & \ldots  & $-$5.71 \\
J1233+1602   & \ldots  & $-$5.31 & $-$5.24 \\
J1234$-$0228 & \ldots  & \ldots  & $-$6.81 \\
J1238+1946   & $-$0.67 & $-$6.12 & $-$6.02 \\
J1538+0252   & \ldots  & \ldots  & $-$7.18 \\
J1625+3632   & $-$3.02 & \ldots  & $-$4.39 \\
J2103$-$0027 & \ldots  & \ldots  & $-$6.95 \\
J2119$-$0018 & \ldots  & \ldots  & $-$7.28 \\
J2132+0754   & \ldots  & \ldots  & $-$6.12 \\
J2228+3623   & \ldots  & $-$4.71 & $-$6.52 \\
\noalign{\smallskip}
\hline
\end{tabular}
\label{tab:Z}
\end{center}
\end{table}

The results of our Balmer line fits using our pure hydrogen grid are
displayed in Figure \ref{fg:DA1}. For the five ELM WDs which also have
helium lines in their optical spectra, we proceed using the exact same
approach detailed above but in addition to the Balmer lines, we also
fit two neutral helium lines, namely \heia\ and \heib\ in order to
measure the helium abundance. Our fits to the these five objects are
displayed in Figure \ref{fg:DAB}.

For the fits to the \caiih\ H \& \caiik\ K lines and the \mgii\ 
doublet, \Te\ and \logg\ are kept fixed at the values measured from 
the Balmer line fits and only the Ca and Mg abundances are allowed 
to vary. The results of the Ca and Mg line fits are shown in 
Figures \ref{fg:Ca} and \ref{fg:Mg}, respectively.

Finally, the only ELM WD whose fit is not shown in Figures
\ref{fg:DA1} through \ref{fg:Mg} is that of the metal-rich and tidally
distorted ELM WD binary J0745. We refer the reader to
\citet{gianninas14} for a detailed description of the analysis of
J0745.

\subsection{Adopted Physical and Binary Parameters}

We present in Table \ref{tab:par} our adopted atmospheric parameters
for the 61 ELM WDs in our sample based on the spectroscopic analysis
presented in the previous section. In the first column we list the
abbreviated SDSS name for each object, ordered by right ascension,
followed by our spectroscopically determined values of \Te\ and
\logg\ with their associated uncertainties. Our error estimates
combine the statistical error of the model fits, obtained from the
covariance matrix of the fitting algorithm, and the systematic error.

The systematic uncertainties for WDs with \logg\ $\sim$~8.0 have
been estimated by \citet{LBH05} using multiple observations of the
same object and are typically 1.2\% in \Te\ and 0.038 dex in
\logg. However, it is not immediately obvious that these estimates
apply to ELM WDs with \logg\ $\sim$~5--6. We have therefore
performed an analogous analysis to that shown in Figure 8 of
\citet{LBH05} using instead the 13 ELM WDs for which we have
multiple observations -- obtained independently at the MMT and FLWO
using the distinct instrument setups described in Section 2 -- to
calculate the average parameters and standard deviation for each
star. The results of this exercise are displayed in \ref{fg:sigma}
as a function of \Te. The average standard deviation in \logg\ is
0.043~dex. This is quite comparable to the value of 0.038~dex
determined by \citet{LBH05}. On the other hand, we obtain a standard
deviation of 0.43\% in \Te.
  
We display in Figure \ref{fg:EW} the sensitivity of the equivalent
width ($W$) of \hgamma, \hdelta, and \hepsilon\ to variations of
\Te\ (i.e. $dW/d$\Te) as a function of \Te\ for models with
\logg\ =~6.0. We did not consider \hbeta\ since it is not included
in our fits. If we compare this result with that obtained for
\logg\ =~8.0 \citep[see Figure 1 in][]{fontaine03} we first remark
that the temperature where the sensitivity vanishes (i.e. where the
Balmer lines reach their maximum equivalent width and
$dW/d$\Te\ =~0) has shifted from $\approx$13,500~K down to
$\approx$10,500~K. However, to either side of this value, the Balmer
lines are just as sensitive at \logg\ =~6.0 as they are at
\logg\ =~8.0. Therefore, it is not surprising that we achieve a
similar precision.

We must also point out that there is an additional source of
uncertainty in the atmospheric parameters, and all the quantities
derived from them, due to the unknown contribution to the optical
spectrum from the companion. Given that these are all single-lined
systems \citep[unlike SDSS 1257+5428, see][]{kulkarni10} whose
companions are likely more massive, cooler, and hence less luminous
WDs, we expect their contribution to the observed spectrum to be on
the order of a few percent at most. For example, \citet{hermes12c}
estimated that the companion of J0651 contributes $\approx$~4\% of the
flux in the SDSS $g$-band (centered at $\lambda
\approx$~4686~\AA). The contamination is mitigated by the fact that we
restrict our fits to the region blueward of \hbeta\ ($\lambda
<$~4500~\AA) where a cooler WD would only contribute a small fraction
of the observed flux. However, since it is impossible to constrain the
parameters (\Te, \logg) of the unseen companions with our current
data, this remains an additional source of uncertainty. As a
result, we choose to adopt the slightly more conservative
uncertainties from \citet{LBH05}.

Table \ref{tab:par} also lists the stellar mass and stellar
radius of the primary as determined by coupling our \Te\ and
\logg\ determinations with the evolutionary models of
\citet{althaus13} appropriate for low mass He-core WDs. The only
exception is J2345$-$0102 whose \Te\ and \logg\ formally place it
outside the Althaus grid. For this object, we use the evolutionary
models of \citet{panei07} instead.  The formal uncertainty in the
stellar mass is obtained by considering the uncertainties on \Te\ and
\logg\ as well as the uncertainties from the evolutionary models
\citep[see][for a detailed discussion]{althaus13}. However, there
remains sufficient uncertainty in the masses derived from the
\citet{althaus13} models due in large part to the many H shell flashes
predicted for the models with masses in the range
0.18--0.36~\msun. For this reason, we adopt a more conservative
uncertainty of 0.020~\msun\ for all of our mass estimates.

The next column lists $g_{0}$, the extinction corrected SDSS $g$-band
magnitude from Data Release 10 \citep{ahn14} followed by $M_{g}$ the
absolute magnitude determined using the photometric calibrations of
\citet{holberg06}. The second-to-last column combines the apparent and
absolute magnitudes to provide an estimate of the distance in kpc.
Finally, in the last column, we list the cooling age, $\tau_{\rm
  cool}$, which we again infer from the models of
\citet{althaus13}. We note again that the parameters for J0745 are
taken from \citet{gianninas14}.

In Table \ref{tab:bin} we provide the binary parameters based on our
updated atmospheric parameters. First, we list the orbital period,
$P$, and velocity semi-amplitude, $K$, for each system. Based on these
values we compute the mass function of the system. Next, we use the
mass function to compute the minimum companion mass, $M_{2}$, assuming
an orbital inclination angle of $i = 90^{\circ}$, and the most likely
companion mass by taking $i = 60^{\circ}$, except for the three
eclipsing systems discussed below. We also provide the merger times
for those systems that will merge within a Hubble time. The last two
columns provide the orbital separation and the expected gravitational
wave strain. Note that no significant radial velocity variability has
been observed for six ELM WDs in our sample and only upper limits on
their velocity semi-amplitudes are provided. Consequently, we cannot
provide binary parameters for those six systems.

There are three eclipsing systems in our sample: \nltt , J0651, J0751
\citep[][respectively]{kaplan14b,brown11,kilic14b} for which eclipse
modeling provides model independent measurements of the parameters of
the system. In particular, the orbital inclination angle and the mass
of the secondary can be constrained and we adopt these values as
  the most likely secondary mass for these three objects. Both J0651
and J0751 have $i\approx 85^{\circ}$ and it is not surprising that the
secondary masses derived from modeling their eclipses, $M_{2, {\rm
    eclipse}}$~=~0.50~\msun\ and 0.97~\msun, respectively, are in
excellent agreement with the minimum secondary masses of
$M_{2}$~=~0.49~$\pm$~0.02~\msun\ and 0.97~$\pm$~0.06~\msun,
respectively, computed assuming $i$~=~90$^{\circ}$. On the other hand,
\citet{kaplan14b} measured $i$~=~89.67$^{\circ}$~$\pm$~0.12$^{\circ}$
for \nltt\ and obtain $M_{2, {\rm
    eclipse}}$~=~0.72~$\pm$~0.01~\msun\ which is significantly
different from our determination of
$M_{2}$~=~0.81~$\pm$~0.02~\msun. Since our determination of the
companion mass depends on the mass of the primary, via the mass
function, the disagreement between our determination and that of
\citet{kaplan14b} is another symptom of high \logg\ problem
discussed in detail below (see Section \ref{rcomp}).

Finally, Table \ref{tab:Z} summarizes the measured atmospheric
abundances of He, Mg, and Ca. As in \citet{gianninas14}, we adopt
uncertainties of 0.30 dex for all the abundances listed in Table
\ref{tab:Z}. This large uncertainty stems from the fact that we are
only fitting a single Ca, or Mg, line at fixed values of \Te\ and
\logg.

\begin{figure*}[!ht]
\centering
\includegraphics[scale=0.675,angle=-90,bb=62 9 566 784]{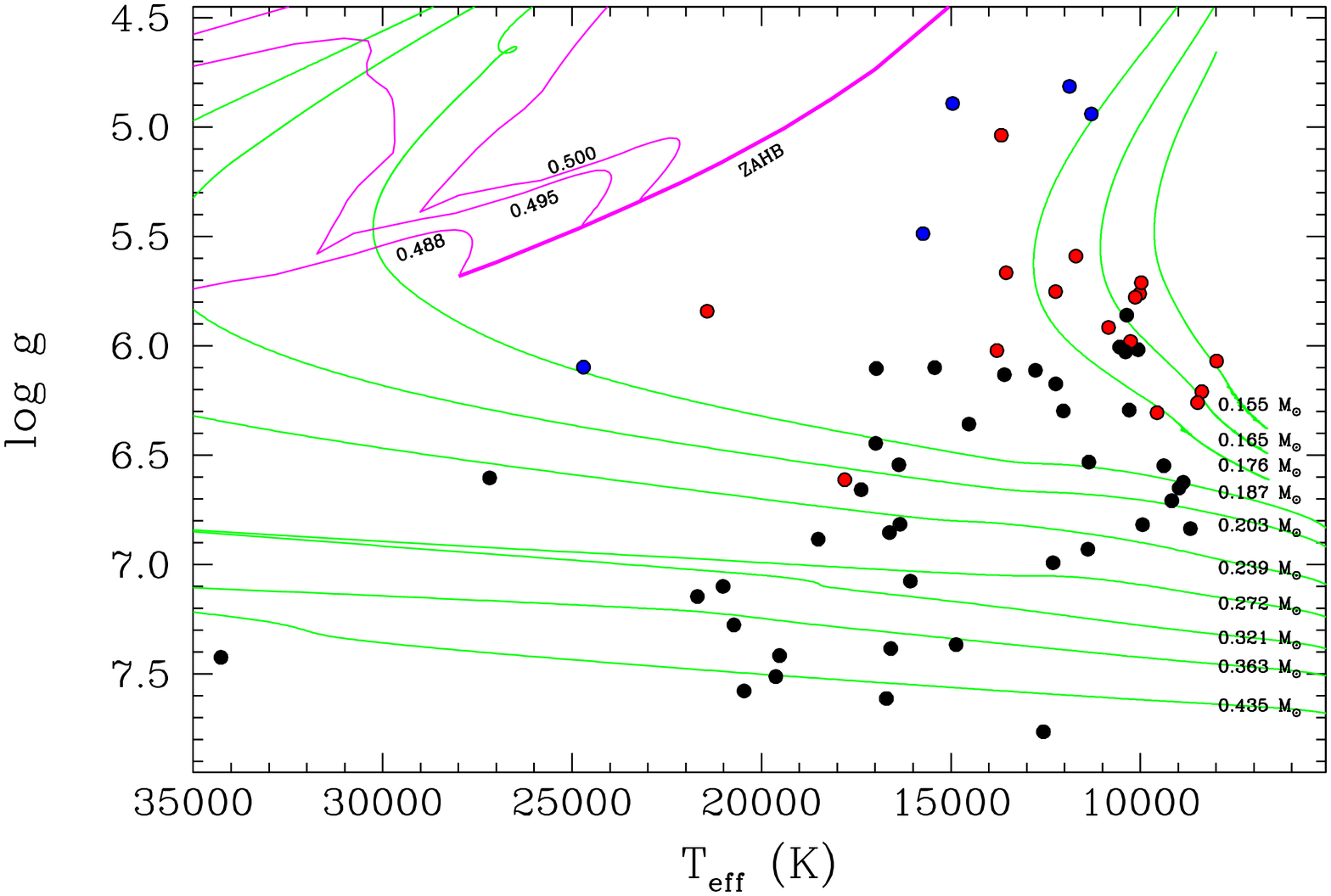}
\figcaption[f08.eps]{The location of the 61 ELM WDs in our
  sample in the \Te-\logg\ plane. The red circles correspond to ELM
  WDs with Ca lines whereas the blue dots denote ELM WDs which have
  both Ca and He lines. The solid green lines correspond to
  theoretical evolutionary tracks for 0.155--0.435~\msun\ He-core WDs
  from \citet{althaus13}. Also shown as the solid magenta lines are
  horizontal branch tracks for 0.488, 0.495 and 0.500 \msun\ stars
  from \citet{dorman93} as well as the zero-age horizontal branch
  (ZAHB) represented by the thick magenta line. \label{fg:tg}}
\end{figure*}

In Figure \ref{fg:comp}, we show a comparison of the atmospheric
parameters from this work compared to those published in the previous
ELM survey papers. In the bottom panel, we see that \Te\ matches
particularly well for lower values, while previous analyses
yield somewhat lower \Te\ at higher values. A somewhat more
pronounced difference is noticeable in the comparison of surface
gravities. The new \logg\ values we obtained are almost systematically
higher than those from previous work. Both of these trends are almost
certainly due to the fact that the models used in previous analyses
did not included the new Stark broadening tables from
\citet{TB09}. The inclusion of these new calculations has already been
shown to produce exactly the same systematic shift of the atmospheric
parameters when compared to the previous generation of models
\citep[see Figure 9 of][]{gianninas11}.

\section{Discussion}

\subsection{Metals in ELM WDs}

We plot in Figure \ref{fg:tg} our entire sample of ELM WDs in the
\Te-\logg\ plane. As a guide, we also plot the evolutionary tracks for
He-core WDs from \citet{althaus13}. Note that, in the interest of
clarity, only the final cooling portion of the Althaus tracks are
shown; we omit the H-shell flashes for models between 0.187 and
0.363~\msun. We also plot horizontal branch tracks with
[Fe/H]~=~$-$1.48 from \citet{dorman93} for 0.488, 0.495, and
0.500~\msun\ stars as well as the zero-age horizontal branch. We can
see that the WDs from the ELM Survey lie in a region where we only
expect He-core WDs to be found.  Figure \ref{fg:tg} also reveals that
nearly all the ELM WDs with \logg~$<$~6.0 are polluted with Ca. The
only exception to this rule is J0818+3536.  However, the spectrum of
J0818+3536 has a S/N~=~17 (see the second panel of Figure
\ref{fg:DA1}). The noise level in the continuum between \hepsilon\ and
H8 could easily conceal a weak Ca line. It seems therefore that the
presence of metals in the lowest mass ELM WDs is a rather ubiquitous
phenomenon and possibly linked to the evolution of these objects.

\begin{figure}[!t]
\includegraphics[scale=0.535,bb=45 17 592 759]{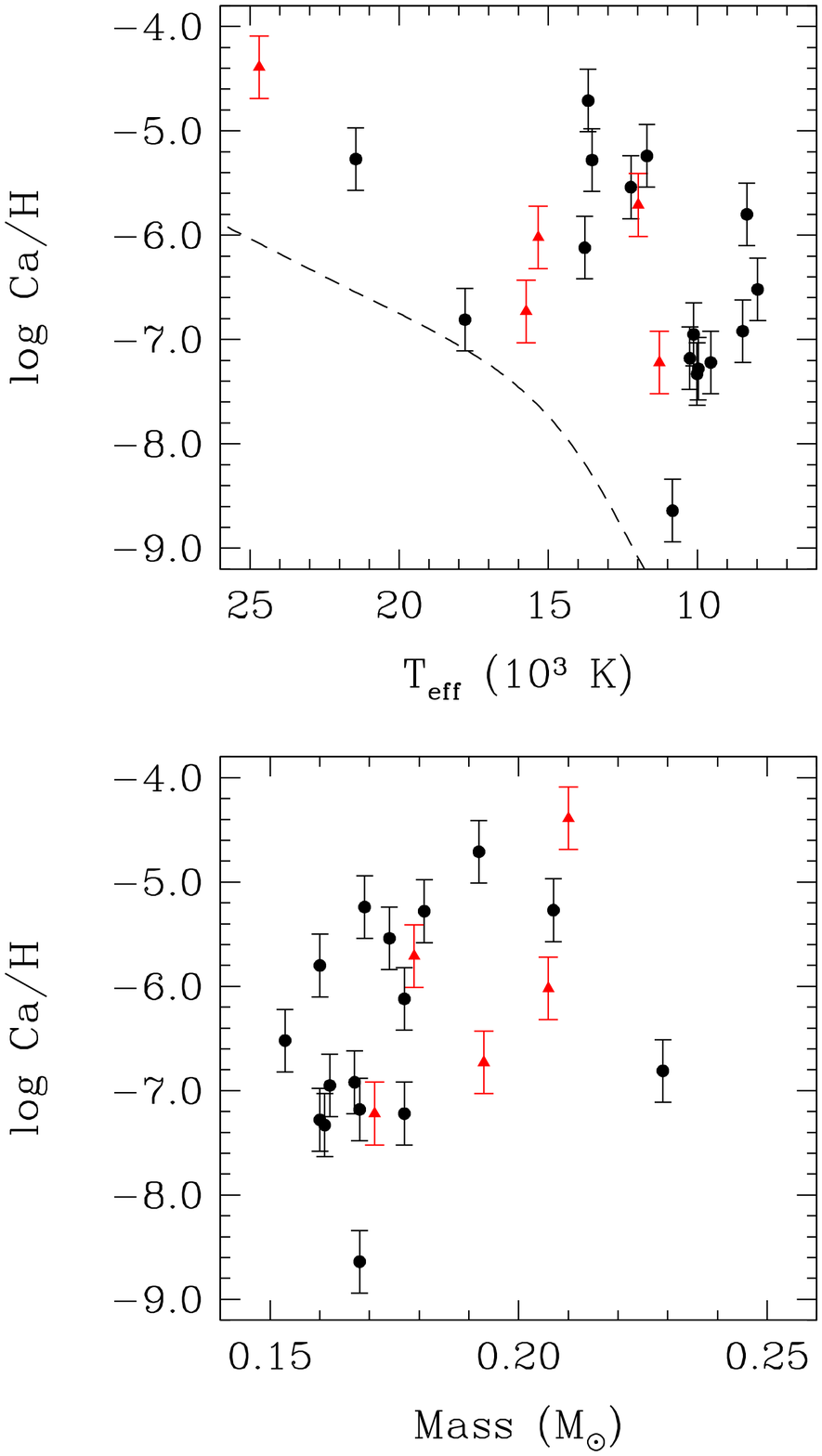}
\figcaption[f09.eps]{Measured Ca abundances plotted as a function of
  \Te\ (top) and stellar mass (bottom). Red triangles represent ELM
  WDs which also contain He. The dashed line corresponds to the
  detection threshold for Ca lines in our spectra. \label{fg:plotCa}}
\end{figure}

\begin{figure}[!ht]
\includegraphics[scale=0.475,bb=20 157 592 609]{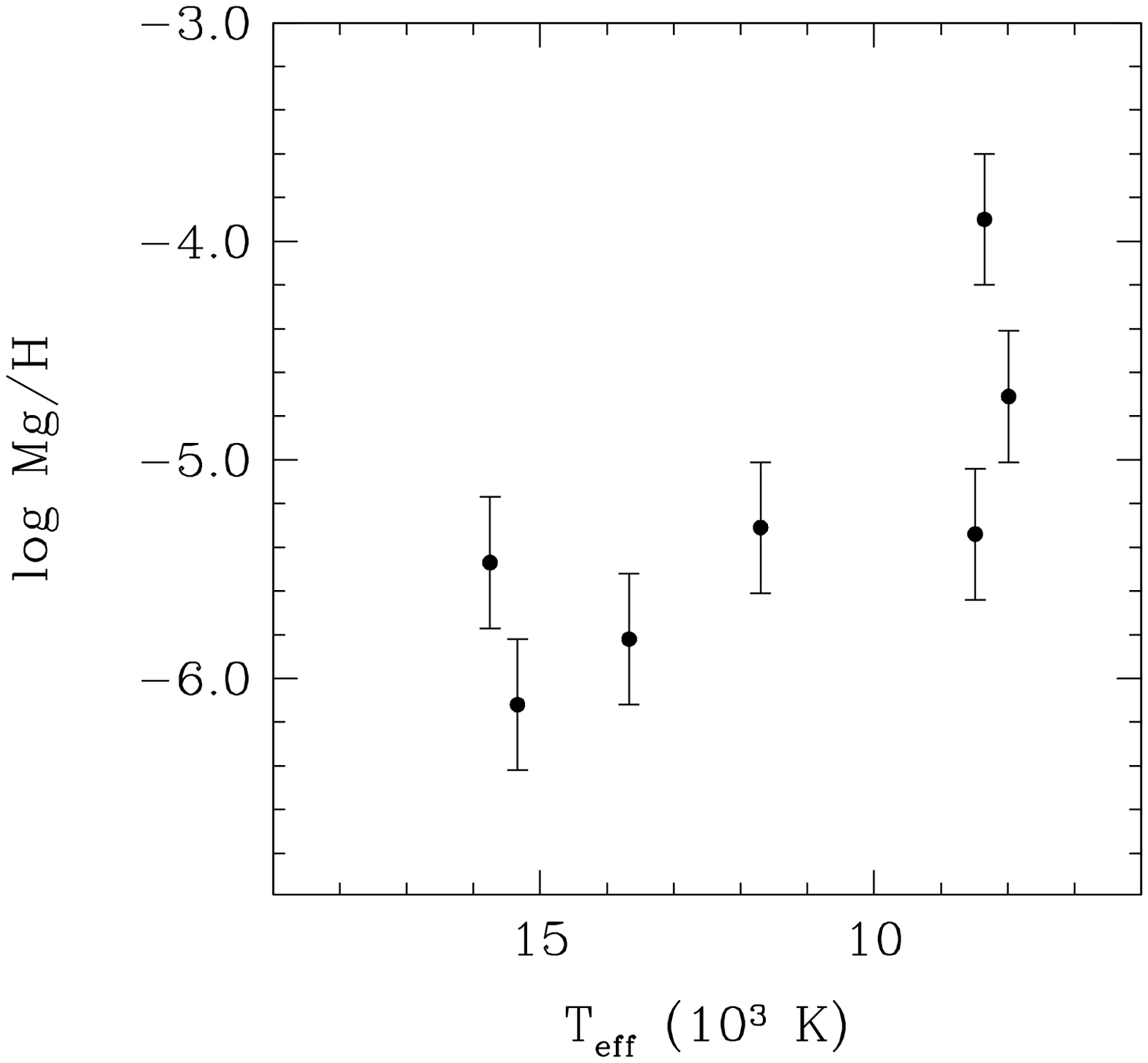}
\figcaption[f10.eps]{Measured Mg abundances plotted as a function of
  \Te.
\label{fg:plotMg}}
\end{figure}

In Figure \ref{fg:plotCa}, we plot the measured Ca abundances as a
function of both \Te\ (top) and the stellar mass (bottom). ELM WDs
which also contain helium are shown in red. \citet{kaplan13} suggested
that H-shell flashes may mix the interior of the WD subsequently
bringing metals back to the surface. We would then expect that as the
WD cools and shell flashes cease, the metals would simply diffuse out
of the atmosphere leading to increasingly lower abundances as a
function of \Te. However, the overall distribution of Ca
abundances in the upper panel of Figure \ref{fg:plotCa} does not
display any obvious trend. A simple linear fit to the data yields a
$p$-value of 0.011. A similar analysis of the data in the lower
panel of Figure \ref{fg:plotCa} gives $p$~=~0.069.  These results
indicate that the Ca abundance distributions are not strongly
correlated to either \Te\ or the mass of the WD. Finally, we also
show in Figure \ref{fg:plotCa} an estimate of the detection limit
for Ca.  Taking into account the typical S/N and resolution of our
spectra, we estimate a minimum equivalent width of 40~m\AA\ for Ca
lines to be detectable. Our measured Ca abundances are consistent
with this detection limit.

The distribution of Mg abundances as a function of \Te\ is
plotted in Figure \ref{fg:plotMg}. A linear fit in this case
produces $p$~=~0.046, once again indicating that there is no strong
correlation. In any case, with only seven Mg detections, as well as
the large uncertainties, it would be difficult to draw any
meaningful conclusions.

The shell flash scenario is also at odds with the fact that
metals are observed in ELM WDs where the evolutionary models do not
predict shell flashes (i.e. for $M <$~0.18~\msun). This
inconsistency suggests that the evolutionary models are possibly in
error. On the other hand, since we have adopted an uncertainty of
0.02~\msun\ for our mass estimates, it's entirely possible that the
true mass of these objects places them in the regime where shell
flashes do indeed occur.

Another difficulty with the shell flash scenario is that the diffusion
timescales for metals in the atmospheres of WDs is typically much
shorter than the evolutionary timescale \citep{paquette86,koester06},
even in ELM WDs \citep{hermes14a}. Even if shell flashes are the
mechanism bringing metals to the surface, some other physical process
must be working to keep them in the atmosphere.  It is possible that
radiative levitation could act against gravitational settling
\citep{chayer10,dupuis10,chayer14}. It is not yet clear if the
considerably lower surface gravity in ELM WDs would allow radiative
levitation to support metals in the atmosphere.  Detailed calculations
of radiative levitation in ELM WDs will need to be performed to
explore this possibility \citep{hermes14a}.

In canonical mass WDs, the presence of metals has been
successfully shown to be the result of ongoing accretion from
circumstellar debris disks formed by the tidal disruption of planetary
bodies 
\citep{debes02,farihi10a,farihi10b,jura03,jura06,jura08,jura07,melis10}. These are detected through excess flux in the IR
\citep{jura03,kilic06,farihi09,barber12}. In the case of ELM
WDs, due to the close nature of these binary systems, we would expect
that any disks would in fact be circumbinary. However, the
formation of such circumbinary disks is problematic. The Roche radius 
of a WD is typically $\sim$~1.0--1.5~\rsun\ 
\citep{jura03,vh07,rafikov11b}. The critical radius for stable orbits 
around a circularized binary is $\sim$2 times the orbital separation 
\citep{holman99}. As it turns out, of the 20 metal-rich ELM WDs, 
only J0745 has and orbital separation (0.63~$\pm$~0.06~\rsun) small 
enough to allow for an orbiting body to pass within its Roche radius. 
There is currently no evidence for debris disks around ELM WDs and the 
orbital separations for virtually all the metal-rich WDs rules out the 
debris disk scenario to explain the presence of metals in ELM WDs.

The ongoing efforts of the ELM Survey will be crucial in increasing
the sample size of these metal-rich ELM WDs. In addition, detailed
analyses based on high-resolution spectroscopy of metal-rich ELM WDs
would allow for more accurate abundance measurements. If multiple
metal lines are detected, and three parameter fits are performed
(i.e. \Te, \logg, and abundances), this would greatly increase the
precision and accuracy of the measured abundances. Until such
observations and studies are performed, it will remain difficult to
draw any firm conclusions regarding the origin of metals in the
atmospheres of ELM WDs.

\subsection{Helium in ELM WDs}

In addition to Ca and Mg, five ELM WDs have optical spectra where
we observe He lines. It is interesting to note that with the
exception of J1625+3632 at \Te\ $\approx$~25,000~K, the remaining
four ELM WDs are clustered together at low \logg\ in Figure
\ref{fg:tg}.  These are also the four WDs with the highest measured
He abundances in our sample. Indeed, the measured He abundances (see
Table \ref{tab:Z}) for J1141+3850, J1157+0546, and J1239+1946 are
unusually high with \loghe\ $\approx$~$-0.5$. The ELM WD companion
to PSR~J1816+4510 \citep{kaplan13} with \Te\ =~16,000~$\pm$~500~K,
\logg\ =~4.9~$\pm$~0.3, and \loghe\ =~0.0~$\pm$~0.5 is another
example of an ELM WD with very similar atmospheric
parameters. \citet{althaus13} report that the predicted shell
flashes ``markedly reduce the hydrogen content of the star'' through
a rapid and intense episode of CNO burning, producing He in the
process. We postulate that the presence of important quantities of
He, not metals, could very well be the signature of a recent shell
flash.

\begin{figure}[!t]
\includegraphics[scale=0.435,bb=20 107 572 659]{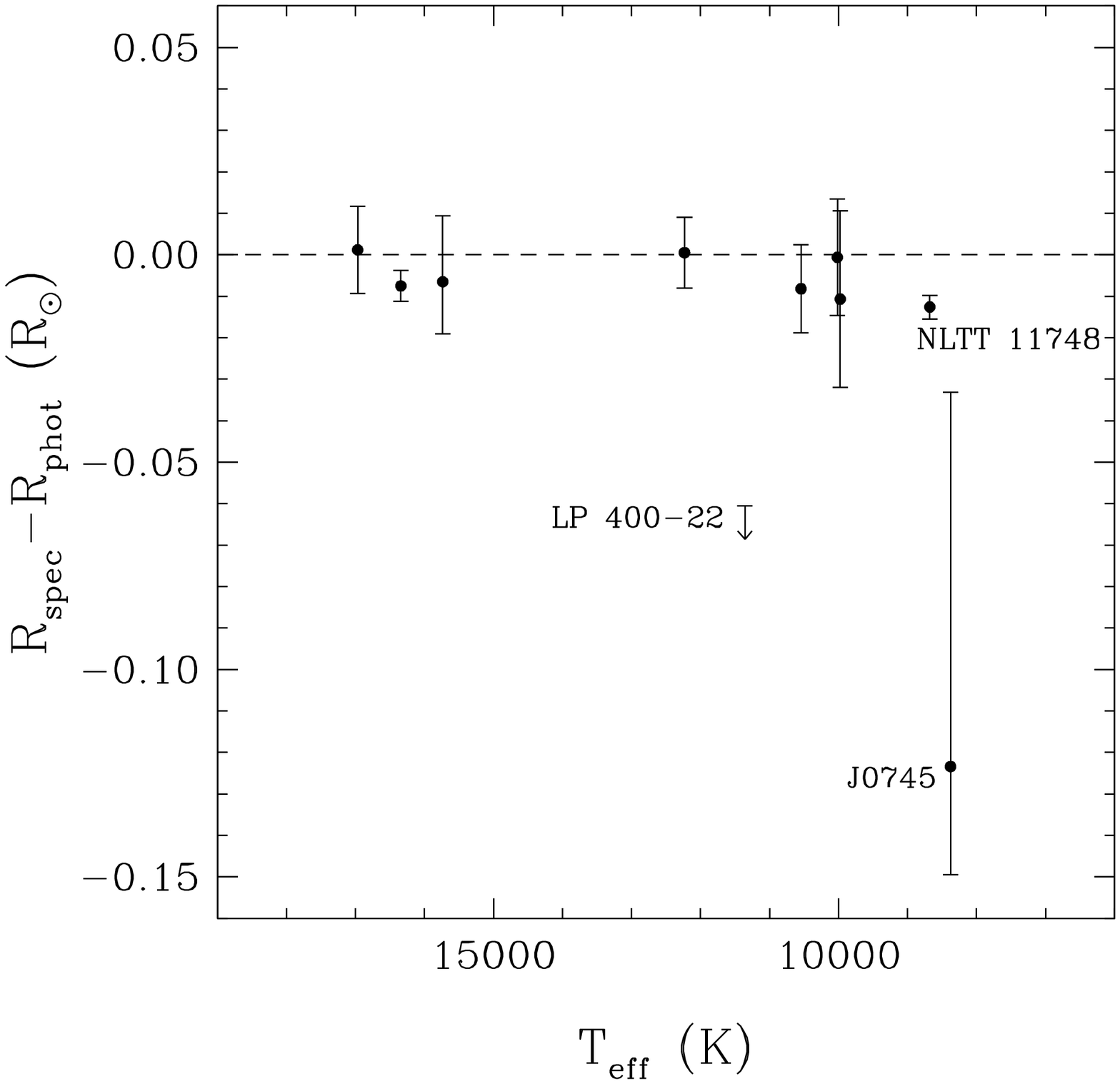}
\figcaption[f11.eps]{Differences in stellar radii as determined from
  our spectroscopic analysis ($R_{spec}$) and from light curve
  analyses ($R_{phot}$), in units of \rsun, plotted as a function of
  \Te. The error bars represent the errors of the two independent
  radius measurements combined in quadrature. As a guide, we plot as a
  dotted line the locus where both the independent radius measurements
  are equal. \label{fg:rad}}
\end{figure}

\subsection{Radius Comparison}\label{rcomp}

Model independent measurements of stellar parameters are crucial for
testing the validity of theoretical models. The nature of several ELM
WD binaries afford us just such a possibility. Ellipsoidal variations
due to tidal distortions and eclipse modeling provide
model-independent measurements of the stellar radius. There are three
eclipsing ELM WDs in the sample analyzed here, J0651, J0751, and
\nltt\ \citep[][respectively]{hermes12c,kilic14b,kaplan14b}. Several
other eclipsing ELM WDs have been discovered as well including
GALEX~J1717+6757 \citep{vennes11} and CSS~41177
\citep{bours14}. Furthermore, there are eight ELM WDs which show
photometric variability attributed to ellipsoidal variations. These
include J0745 \citep{gianninas14} and seven more systems analyzed in
\citet{hermes14a}. Note that J0651 and J0751 both display ellipsoidal
variations and eclipses in their observed light curves.  Finally, the
ELM WD \lp, a unique system that is leaving the galaxy, has a measured
parallax which also constrains the radius of the star
\citep{kilic13b}.

\begin{figure}[!t]
\includegraphics[scale=0.465,bb=25 117 592 659]{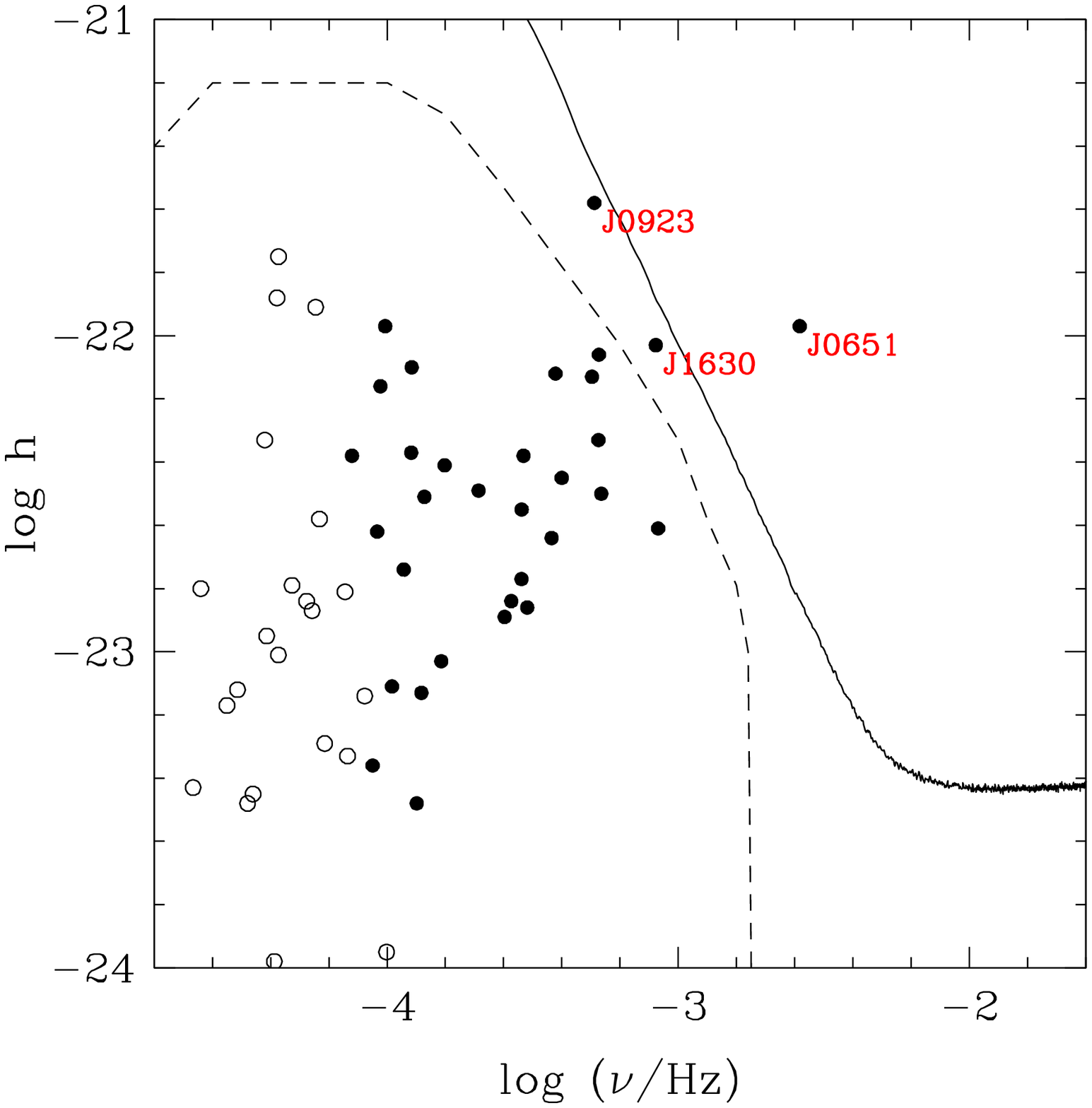}
\figcaption[f12.eps]{Gravitational wave strain versus frequency for
  our ELM WD binaries. Filled circles represent merger systems whereas
  open circles correspond to non-merger systems. The solid line
  represents the new sensitivity curve for of $eLISA$ after two years
  of observation \citep{amaro13}. The dashed line shows the predicted
  average Galactic foreground after one year of integration 
  \citep{nelemans01}. The three labeled objects are the ELM WD binaries 
  which could be verification sources for $eLISA$. \label{fg:GW}}
\end{figure}

In Figure \ref{fg:rad} we plot the difference in the determinations of
the stellar radii for ELM WDs derived from our spectroscopic analysis
and from the model-independent determinations enumerated above, as a
function of \Te. We see that the agreement between the
spectroscopically inferred radii and the model-independent values is
quite good for \Te~$>$~10,000~K, given the uncertainties. However,
there are three objects where the radius estimates do not agree:
\nltt, \lp, and J0745. In all three cases, the spectroscopic
determination underestimates the radius. The result for \nltt\ in
particular is 4$\sigma$ significance. In the case of J0745, it is
possible that the spectroscopic mass and radius determinations
suffer from the assumption that the WD is on its terminal cooling
track \citep[see][for a detailed discussion]{hermes14a}. As for 
LP~400-22, its measured parallax implies a lower limit of
$R$~=~0.099~\rsun\ \citep[see][for a detailed discussion]{kilic13b}.

Underestimating the radius is analogous to overestimating the mass or
the surface gravity since WDs have an inverse mass-radius
relationship. This is most likely a manifestation of the ``high
\logg\ problem'' \citep{kepler07,tremblay10,gianninas11} but in the
regime of ELM WDs, as first noted in \citet{gianninas14}. This well
documented phenomenon is a consequence of the 1D treatment of
convection via the mixing-length theory \citep{tremblay11}, the same
treatment currently implemented in the model grids used for our
analysis. \citet{tremblay13a} presented a new series of models which
employ a 3D hydrodynamical treatment of convection. \citet{tremblay13b} 
then showed how these models effectively solve the high \logg\ problem 
and derive corrections which can be applied to atmospheric parameters 
determined from 1D models. Figure 4 in \citet{tremblay13b} shows these 
corrections which systematically imply lower \logg\ values than those 
determined from 1D models. Unfortunately the model grid in 
\citet{tremblay13b} only extends to \logg\ =~7.0 and therefore does not 
cover parameters appropriate for ELM WDs. However, we remark that the 
corrections for the models at \logg\ =~7.0 are greatest for
10,000~K~$>$~\Te\ $>$~8000~K. This roughly matches the range in
\Te\ where we see the largest discrepancy in our radius
determinations. With only seven of our targets with
  \Te\ $<$~9500~K, only 11\% of our sample is actually affected. We
(in collaboration with P.-E. Tremblay) are currently computing 3D
models appropriate for ELM WDs to resolve this issue.

\subsection{Gravitational Waves}

Using our determinations of the orbital period, distance and the
masses of both components in our ELM WD binaries (assuming an
inclination of $i=60^{\circ}$), we can calculate the gravitational
wave strain, $h$, expected from these systems \citep[see][and
  references therein]{roelofs07}. The results of these calculations
are listed in the last column of Table \ref{tab:bin}. We plot in
Figure \ref{fg:GW} $\log h$ as a function of $\log \nu$, where $\nu =
2/P_{\rm orb}$ in Hz, for each system. The majority of our ELM WD
binaries are contained within the region in Figure \ref{fg:GW}
characterizing the average galactic foreground emission
predicted from population synthesis models \citep{nelemans01}. Our
results suggest that an important fraction of the the galactic
foreground at mHz frequencies is due to short-period ELM WD
binaries. J0651 is the only ELM WD system that would be a
verification source based on past $LISA$ sensitivity curves
\citep{larson00}. Figure \ref{fg:GW} shows that the new sensitivity
curve for the revised $eLISA$ mission \citep{amaro13} leaves only
three potentially viable sources. Note that the sensitivity curve
shown here is expressed in terms of the dimensionless strain $h$ for
monochromatic (periodic) sources for an integration time, $T$, of
two years, i.e. the dimensionless value $\sqrt{S(f)/T}$, where
$S(f)$ is the $eLISA$ equivalent-strain noise and $f$ is the
frequency. However, the only real verification source remains
J0651 as it lies well within the projected sensitivity region for
$eLISA$. Given enough observation time and a high enough S/N, J0923
and J1630 could potentially be detected as well. As the ELM Survey is
ongoing, there also remains a strong possibility that more
verification binaries could be uncovered \citep[e.g. ][]{kilic14a}.

\begin{figure}[!t]
\includegraphics[scale=0.425,bb=20 77 592 679]{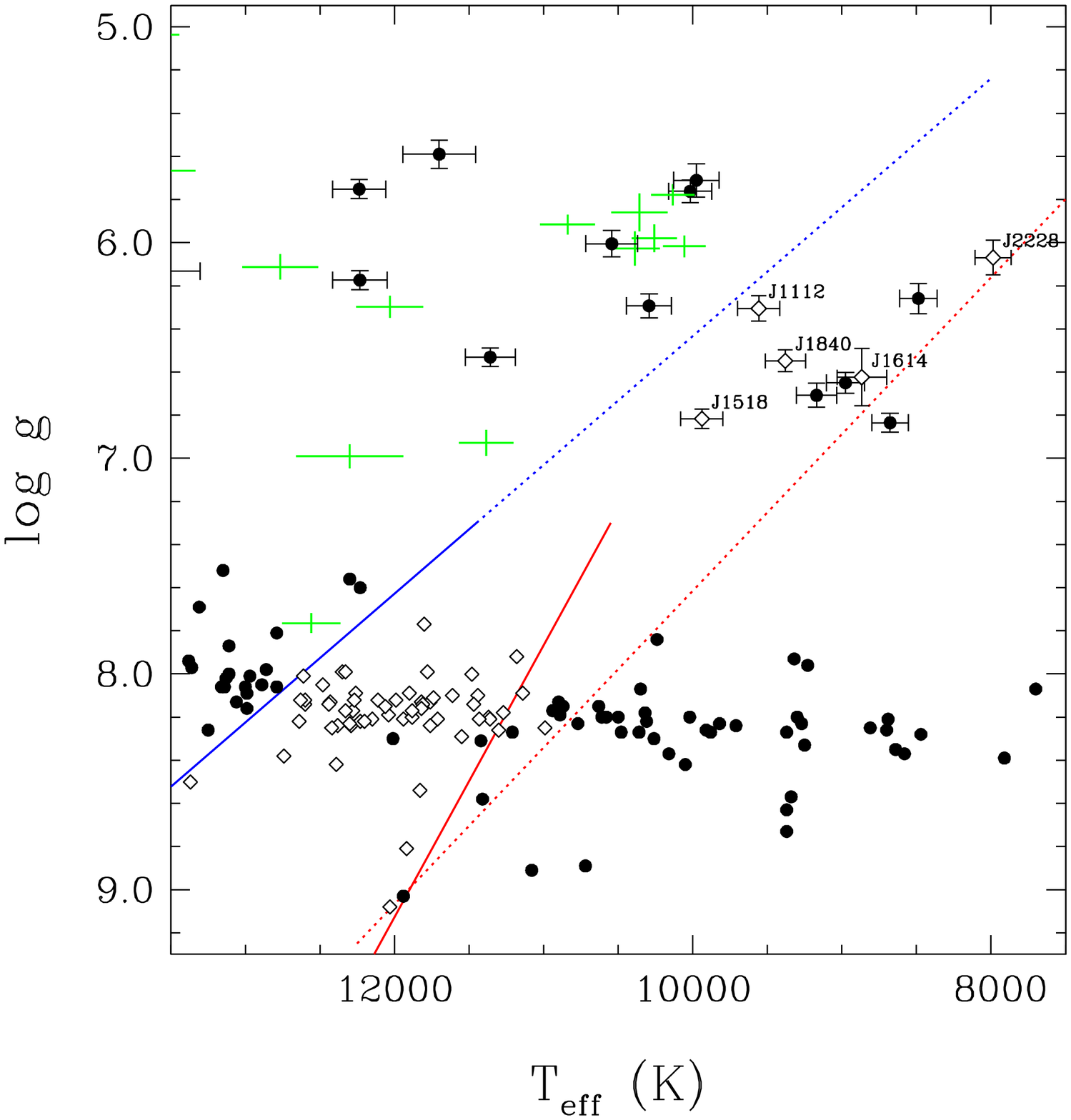}
\figcaption[f13.eps]{Region of the \Te -\logg\ plane containing the ZZ
  Ceti instability strip (lower left) and the five currently known ELM
  pulsators, which are labeled in the figure. Pulsators are identified
  as white diamonds whereas WDs which have been confirmed as
  photometrically constant are represented as black dots. The blue and
  red lines represent the empirical boundaries of the ZZ Ceti
  instability as determined by \citet{gianninas11}. The dotted lines
  denote tentative boundaries which match the location of both
  instability strips. The green errorbars denote the remaining ELM WDs
  from our sample which have not yet been investigated for photometric
  variability. \label{fg:ZZ}}
\end{figure}

\subsection{Instability Strip}

In this section, we use our new atmospheric parameter determinations
to update the current view of the ELM WD instability strip for He-core
pulsators and compare it with the most recent determination of the ZZ
Ceti instability strip populated by CO-core WDs. In order to be able
to make this comparison in a self-consistent manner, we plot in
Figure~\ref{fg:ZZ} only WDs which have been analyzed using the exact
same fitting technique used in the present analysis and using model
atmospheres which employ the same parametrization of the mixing length
theory to model convection
\citep[i.e. ML2/$\alpha$~=~0.8][]{tremblay10}.  For this reason, some
of the WDs shown in Figure~5 of \citet{hermes13b} are not included
here. For analogous reasons, we choose not to plot in
Figure~\ref{fg:ZZ} the theoretically predicted boundaries of the
extended ZZ Ceti instability computed by \citet{vg13} since their
pulsation models assume a parametrization of the mixing length theory
equivalent to ML2/$\alpha$~=~0.6 in the atmosphere.

The lower portion of Figure \ref{fg:ZZ} includes the 56 pulsating ZZ
Ceti WDs as well as the 145 photometrically constant DA WDs from
\citet{gianninas11}. We also include GD~518, recently discovered to be
the most massive known pulsating ZZ Ceti WD \citep{hermes13a}. The ELM
WDs analyzed in this paper are represented in Figure \ref{fg:ZZ} by
analogous symbols with error bars. We plot the current sample of five
ELM WD pulsators as well as non-variables listed in
\citet{hermes12b,hermes13c,hermes13b}. For the purposes of empirically
defining the instability strip of ELM WDs, we consider the 20 ELM WDs
analyzed by \citet{hermes14a} as non-variables from the point of view
of pulsations. Their photometric variability is perfectly consistent
with ellipsoidal variations and the observed periods correlate almost
perfectly with the orbital periods. We also plot a number of
additional non-variables ELM WDs from
\citet{steinfadt10,steinfadt12}. Finally, ELM WDs which have not been
observed for photometric variability are plotted as green error bars.
Figure \ref{fg:ZZ} clearly shows that the situation for ELM WD
pulsators is not nearly as clear cut as it is for their more massive
counterparts in the ZZ Ceti instability whose blue and red edges are
fairly well constrained. However, if one extrapolates the empirical
blue edge of the ZZ Ceti instability strip to lower \logg, the ELM WD
pulsators do conform to that same boundary. The same does not apply to
the empirical red edge. Indeed, the red edge would need to be
essentially parallel to the blue edge in order to match the location
of both instability strips. It is interesting to note that the
theoretical boundaries predicted by \citet{vg13} are qualitatively
similar in this regard. It is obvious that any attempt to map out the
instability strip of the coolest, and least massive, class of
pulsating WDs will require identifying many more pulsators than the
five that are currently known.  Identifying new ELM WD pulsators is
important since asteroseismological studies of these stars will reveal
the details of their internal structure leading to a better
understanding of the evolution of ELM WDs.

\section{Conclusions}

We have performed a homogeneous spectroscopic analysis of the entire
sample of ELM WDs from the ELM Survey using the latest model
atmosphere grids appropriate for these stars. We provide updated
atmospheric and binary parameters for 61 ELM WDs binaries. In
particular, we note that nine ELM WDs have minimum secondary masses of
$M_{2}$~$>$~0.80~\msun\ and six systems have 0.70~\msun\ $< M_{2}
<$~0.80~\msun. Among these 15 binaries, seven will merge within a
Hubble time and thus represent likely progenitors of underluminous .Ia
supernovae, as postulated by \citet{bildsten07} and \citet{shen09}.
For the first time, we also provide systematic measurements of the
atmospheric abundances of He, Mg and Ca. Unfortunately, the
distributions of Ca and Mg as a function of \Te\ and mass do not
yield any clues as to the origin of the metals. Furthermore, shell
flashes cannot explain the presence of metals in the least massive ELM
WDs. Conversely, the detection of He in ELM WDs may be the
signpost that a shell flash has recently occurred. It is also
unlikely that metal-rich ELM WDs harbor debris disks formed from the
tidal disruption of planetary bodies like their more massive
counterparts. The orbital separations are simply too large to allow
a rocky body to venture within the tidal radius of the WD.

We have also shown that stellar radii derived from our spectroscopic
fits do not agree with radii from model-independent measurements
for \Te\ $<$~10,000~K, the likely consequence of the 1D treatment of
convection in our model atmospheres. Our results also indicate that
ELM WD binaries possibly comprise an important fraction of the
galactic gravitational wave foreground emission while the
shortest-period system, J0651, represents a strong verification source
for eventual gravitational wave detectors like $eLISA$. Finally, we
showed that the current state of the instability strip of ELM WD
pulsators is not nearly as obvious as that of the ZZ Ceti instability
strip. Many more ELM WD pulsators will need to be identified if we are
to map out the instability strip of pulsating He-core WDs in an
analogous manner. The one overarching theme is that we must continue
the search for ELM WDs if we are to understand the origin, evolution,
and ultimate fate of these most intriguing,and extreme, products of
binary evolution.\\ \\

We would like to thank both referees for a careful reading our
manuscript and for their numerous suggestions that helped to improve
this paper. We gratefully acknowledge the support from the NSF and
NASA under grants AST-1312678 and NNX14AF65G, respectively. This
research makes use of the SAO/NASA Astrophysics Data System
Bibliographic Service. This project makes use of data products from
the Sloan Digital Sky Survey, which has been funded by the Alfred
P. Sloan Foundation, the Participating Institutions, the National
Science Foundation, and the U.S. Department of Energy Office of
Science.  This work was supported in part by the Smithsonian
Institution. This work is funded in part by the NSERC Canada and by
the Fund FRQ-NT (Qu\'ebec).

{\it Facilities:} MMT (Blue Channel Spectrograph), FLWO:1.5m (FAST)

\bibliographystyle{apj}
\bibliography{biblio}

\end{document}